\documentclass[preprint2]{aastex}
\newcommand{\etal}{{et al.}}
\newcommand{\h}{$^{\rm h}$}
\newcommand{\m}{$^{\rm m}$}
\newcommand{\s}{$^{\rm s}$}
\newcommand{\lsun}{L$_{\odot}$}

\begin{document}

\title{The Stellar Content of M31's Bulge\altaffilmark{1}}

\author{Andrew W. Stephens\altaffilmark{2}}
\affil{Pontificia Universidad Cat\'{o}lica de Chile, 
Departamento de Astronom\'{\i}a y Astrof\'{\i}sica, 
Cassilla 306, Santiago 22, Chile; 
and Princeton University Observatory, 
Peyton Hall - Ivy Lane, 
Princeton, NJ 08544-1001; stephens@astro.puc.cl}

\author{Jay A. Frogel \altaffilmark{3,4}}
\affil{NASA Headquarters, 300 E Street SW, Washington, DC  20546}

\author{D. L. DePoy}
\affil{The Ohio State University, Department of Astronomy,
140 West 18th Avenue, Columbus, OH  43210}

\author{Wendy Freedman}
\affil{Carnegie Observatories}

\author{Carme Gallart \altaffilmark{5}}
\affil{Instituto de Astrof\'{\i}sica de Canarias, 
38200 Tenerife, Canary Islands, Spain}

\author{Pascale Jablonka}
\affil{Observatoire de Paris-Meudon}

\author{Alvio Renzini}
\affil{European Southern Observatory}

\author{R. Michael Rich}
\affil{University of California at Los Angeles}

\and 

\author{Roger Davies}
\affil{University of Durham}

\altaffiltext{1}{Based on observations with the NASA/ESA Hubble Space 
Telescope obtained at the Space Telescope Science Institute, which is
operated by AURA for NASA under contract NAS5-26555.}

\altaffiltext{2}{Princeton-Catolica Prize Fellow}

\altaffiltext{3}{Permanent address: The Ohio State University, 
Department of Astronomy, 140 West 18th Avenue, Columbus, OH  43210}

\altaffiltext{4}{Visiting Investigator, 
Department of Terrestrial Magnetism, 
Carnegie Institution of Washington.}

\altaffiltext{5}{Ramon y Cajal Fellow}

\begin{abstract}

In this paper we analyze the stellar populations present in M31 using
nine sets of adjacent HST-NICMOS Camera 1 and 2 fields with
galactocentric distances ranging from $2'$ to $20'$. These infrared
observations provide some of the highest spatial resolution measurements
of M31 to date; our data place tight constraints on the maximum
luminosities of stars in the bulge of M31.  The tip of the red giant
branch is clearly visible at $M_{bol} \sim -3.8$, and the tip of the
asymptotic giant branch (AGB) extends to $M_{bol} \sim -5$.  This AGB
peak luminosity is significantly fainter than previously claimed;
through direct comparisons and simulations we show that previous
measurements were affected by image blending.  We do observe
field-to-field variations in the luminosity functions, but simulations
show that these differences can be produced by blending in the higher
surface brightness fields.  We conclude that the red giant branch of the
bulge of M31 is not measurably different from that of the Milky Way's
bulge.  We also find an unusually high number of bright blueish stars
(7.3 arcmin$^{-2}$) which appear to be Galactic foreground stars.

\end{abstract}


\section{Introduction} \label{sec:introduction}

The first deep infrared (IR) observations of stars in the bulge of our
Galaxy were carried out by \citet{Frogel1987}.  Measuring the luminosity
function (LF) using the M giant grism surveys of \citet{Blanco1984} and
\citet{Blanco1986}, they found many luminous giants, but noticed that
the LF has a sharp break at $M_{bol} \simeq -4.5$ ($M_K \simeq -7.5$),
with the brightest stars extending to $M_{bol} \simeq -5$.

The tip of red giant branch (RGB), defined by the core mass required for
helium flash, occurs at a luminosity of $M_{bol}\simeq -3.8$.  Any stars
brighter than this limit are therefore on the asymptotic giant branch
(AGB).  The stars observed in the Galactic bulge extend $\sim 1.2$
magnitudes brighter than the tip of the RGB.  Since metal-poor
([Fe/H]$\lesssim -1$) Galactic globular clusters do not exhibit such
luminous AGB stars, this might have suggested a younger age for the
bulge population, since the luminosity of the brightest AGB stars
increases with decreasing age \citep[e.g.][]{Iben1983}.  However
metal-rich globular clusters do have stars which can reach luminosities
of $M_{bol} \simeq -5.0$, while still having ages comparable to the
metal poor clusters \citep{Frogel+Elias1988, Guarnieri1997}.  Moreover,
it has been demonstrated that the stellar population of the Galactic
bulge is dominated by metal-rich stars ([Fe/H]$ \gtrsim -1$)
\citep{McWilliam1994} that are as old as Galactic globular clusters
\citep{Ortolani1995, Feltzing2000, Kuijken2002, Zoccali2002}, and that 
the number of stars brighter than the RGB tip is consistent with the
frequency observed in old, metal rich globular clusters.  In summary,
the stellar population in the bulge of the Milky Way is as old as the
oldest Galactic globular clusters, and old, metal rich stellar
populations are able to produce AGB stars only as bright as $M_{bol}
\simeq -5$.

Work was already underway studying stars in other nearby galaxies to
determine whether the properties of the stars in the Galactic bulge are
typical of all bulges.  The nearest and brightest large spiral, M31 (the
Andromeda Galaxy) was the obvious first choice for a comparison.  Some
of the first measurements of stars in the inner bulge of M31 ($\sim 1$
kpc from the nucleus) were made by \citet{Mould1986}.  He found that
M31's brightest bulge stars were $\sim 1$ magnitude more luminous than
the brightest stars in M31's halo.  \citet{Rich1989} then took spectra
of some of these bright stars, and found most to have properties
characteristic of late-type M giants.

It was unclear what was causing the apparent difference between the
stellar populations of the bulges of the Milky Way and M31.  The
dependence of the AGB peak luminosity on mass (and therefore age and
mass loss) and metallicity pointed to several possible explanations for
this observed difference.  However, all of these explanations implied a
difference in the formation or evolutionary processes of these two
otherwise very similar galaxies.

To see if the luminous stars in M31 are indeed similar to those found in
the bulge of the Milky Way, \citet{Rich1991} measured a sample of $\sim
600$ stars in the inner bulge of M31 $4'$ from the nucleus with the Hale
5m reflector (see \S \ref{sec:rm91}).  Their resulting LF had a drop at
$M_{bol} \sim -4.5$, similar to that seen in BW, but extended to
$M_{bol} \sim -5.5$.  To explain these excess luminous stars they
proposed several theories.  (1) These stars could be younger stars from
the disk superposed on the bulge.  (2) They could be super-metal-rich in
chemical composition, since the luminosity of the brightest AGB stars
increases with metallicity.  (3) The bulge of M31 could have a young
stellar component, since the luminosity of the brightest AGB stars
increases with decreasing age \citep{Iben1983}.  (4) They could be the
result of merged lower-mass main-sequence stars (blue straggler
progenitors), which produces more massive and luminous stars.

Soon thereafter, \citet{Davies1991} imaged the M31 bulge in the near-IR
using the 3.8m UKIRT facility, albeit $7.2'$ from the nucleus, a factor
of two more distant than Rich.  Their LF has an upper limit $\sim 0.5$
mag brighter than that seen in the Galactic bulge.  They argued that
contamination by stars from a young disk, which lies behind the bulge,
is most likely responsible for the luminous stars observed by RM91.

\citet{DePoy1993} carried out a $K$-band survey of $\sim 17000$ stars in 
604 arcmin$^2$ of Baade's Window to check the possibility that the M
giant surveys in the bulge of the MW \citep{Blanco1984, Blanco1986} may
have missed very luminous stars similar to the ones seen in the bulge of
M31.  The \citet{DePoy1993} observations turned up no such population of
luminous stars, and their derived LF is consistent with that obtained by
\citet{Frogel1987}.  In order to compare their Galactic Bulge
observations with those of the bulge of M31, \citet{DePoy1993} rebinned
and smoothed their image to simulate the M31 observations.  The
resulting degraded image showed that few if any of the ``stars'' on this
simulated M31 field corresponded to individual stars on the original
image; most were just random groupings of stars.  A quantitative
analysis showed that the extreme crowding caused an artificial
brightening in the LF of more than one magnitude.  They thus concluded
that the luminous stars seen in M31's bulge were most likely not real,
but an artifact of image crowding.

At nearly the same time \citet{Rich1993} acquired new observations of
five fields in the bulge of M31 with the Palomar IR Imager on the Hale
5m telescope (see \S \ref{sec:rmg93}).  By measuring the LFs in fields with
different expected disk contributions from 2 to 11 arcminutes from the
nucleus, they rejected the hypothesis of \citet{Davies1991} that the
bright stars are disk contaminants.  Using their own model fields, which
showed that they could accurately measure the GB tip despite the
crowding, they also argued against the idea of \citet{DePoy1993} that
the bright stars are stellar blends.  While they did concede that some
of their measurements may have been affected by crowding of up to 1
magnitude, they maintained that they were not generally measuring
clusters of blended images.  Further calculations by \citet{Rich1993}
also showed the numbers of blue straggler progeny stars to be
insufficient to explain the number of luminous stars.

In pursuit of a resolution to this controversy, \citet{Rich1995}
obtained HST Wide-Field Planetary Camera (WFPC1) observations of the
inner bulge of M31.  Regretfully, these observations were taken with
HST's original aberrated optics, reducing the effective resolution to
barely better than was available from the ground.  These observations
also yielded many luminous stars, although not quite as bright as
previously measured.  The data also suggested that the brightest stars
may be concentrated toward the center of M31.

Approaching this problem from the theoretical side, \citet{Renzini1993,
Renzini1998} performed calculations to estimate the number of stars in
all evolutionary stages in each pixel.  He showed that the number of
blends increases quadratically with both the surface brightness of the
target and with the angular resolution of the observations.  Applying
these calculations to existing photometric data for the inner bulge of
M31, he concluded that all previous ground-based observations were
dominated by blends, and even questioned the HST observations of
\citet{Rich1995}, pointing to the measured blue $(R-I)$ colors as
indicative of their blended origin.

With the Wide-Field Planetary Camera 2 (WFPC2), \citet{Jablonka1999}
observed three fields in the bulge of M31 at optical wavelengths.  With
HST's improved resolution, they did not find stars more luminous than
those in the Galactic Bulge, and concluded that previous detections of
very bright stars were likely the result of blended stars due to the
crowding in WFPC1 and ground-based images.

However, since even very luminous evolved stars can go undetected at
optical wavelengths due to molecular blanketing, \citet{Davidge2001}
recently obtained new infrared images of the bulge of M31 with the 3.6m
CFHT.  With the help of adaptive optics (AO), his $JHK$ observations
confirm the optical non-detection of very luminous stars made by
\citet{Jablonka1999}.  Although there is agreement between the brightest
stars measured by \citep{Davidge2001} in the bulge of M31 and the
brightest stars measured in the Galactic Bulge \citep{Frogel1987}, the
luminosity functions still show considerable differences.  The M31 bulge
LF measured by \citet[][Figure 7]{Davidge2001} does {\em not} show a
break at $M_K \sim -7.5$ as is observed in BW, but instead shows a
change in slope at $M_K \sim -8$ and a break at $M_K \sim -8.6$,
indicative of different star formation histories in the MW and M31 if
correct.


Thus it appears that this decade-old controversy has not yet been
completely resolved.  While there now appears to be agreement that the
brightest of the previous measurements were blends, it is still not
certain whether the luminosity functions of the bulges of the Milky Way
and M31 are consistent with one another.  In this paper we will show
that indeed they are, and that even the most recent observations,
including our own, are still affected by blending in the inner regions
of M31.

The layout of the paper is as follows.  We start by describing our
observations in Section \ref{sec:observations}, and our reduction and
photometric techniques in Section \ref{sec:reduction}.  We present the
color magnitude diagrams and luminosity functions in Sections
\ref{sec:cmds} and \ref{sec:lfs} respectively.  Section
\ref{sec:blending} gives a brief theoretical analysis of blending in
M31, followed by detailed simulations of all our fields in Section
\ref{sec:simulations}.  In Section \ref{sec:comparison} we compare our
measurements with previous observations;  \citet{Rich1993} in \S
\ref{sec:rmg93}, \citet{Rich1991} in \S \ref{sec:rm91}, and 
\citet{Davidge2001} in \S \ref{sec:davidge}.  Section \ref{sec:brightstars} 
discusses the bright stars which appear to be Galactic foreground stars,
and we conclude with a brief summary of our results in Section
\ref{sec:conclusions}.

\section{Observations} \label{sec:observations}

\begin{figure*}[htb]
\epsscale{1.5}
\plotone{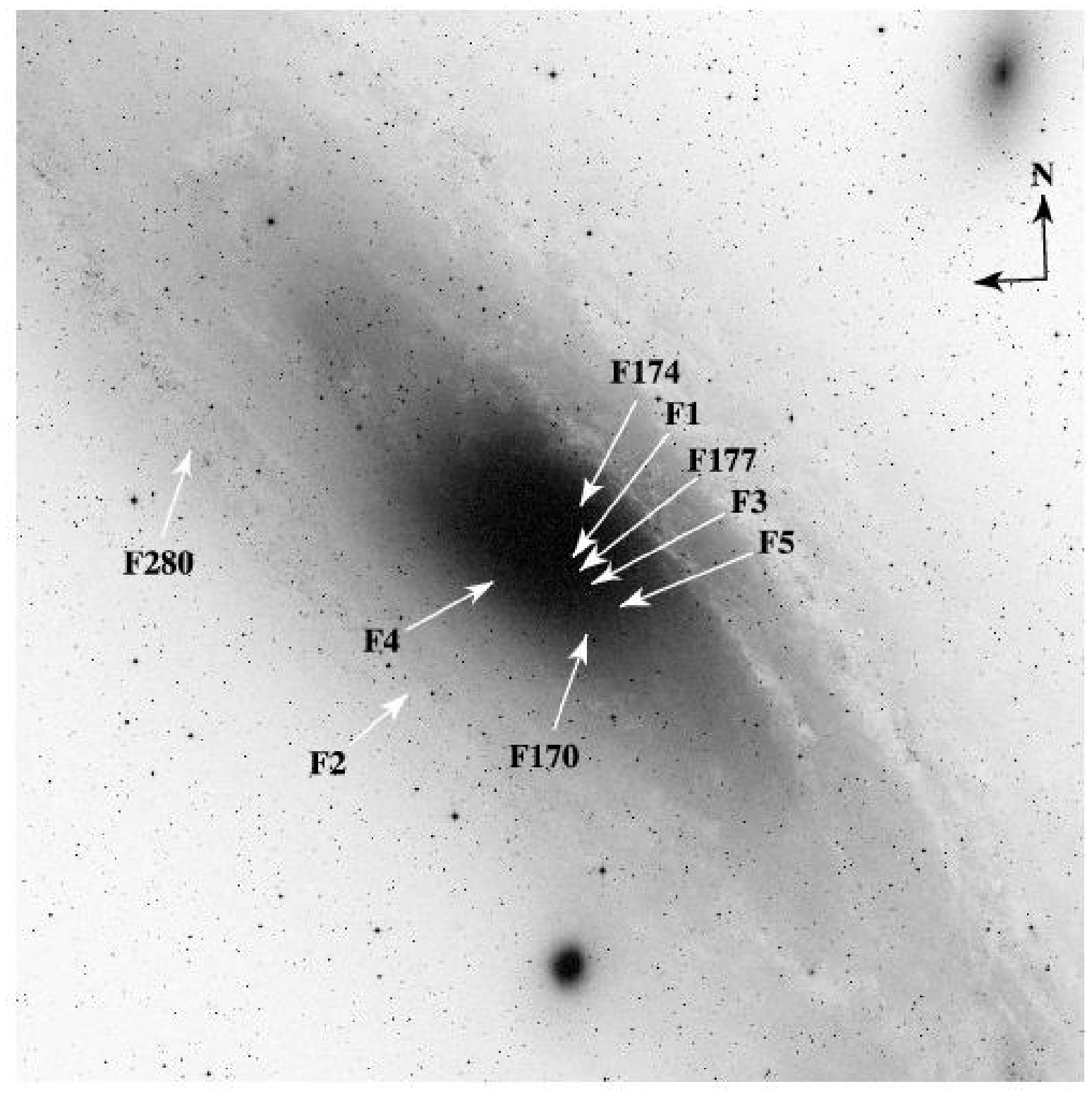} 
\figcaption{
A 1\degr\ field from the Digitized Sky Survey showing the location of the
nine fields used in our analysis.  North is up and East is to the left.
\label{fig:fields}}
\end{figure*}

In this paper we analyze images of M31 taken from two different NICMOS
proposals.  Proposal 7876 imaged the five central fields of
\citet{Rich1993}.  These fields were carefully chosen to sample varying
bulge/disk ratios, and are indicated by F1--F5 on Figure
\ref{fig:fields}.  Proposal 7826 imaged five globular clusters in M31.
The four fields used in this study are indicated on Figure
\ref{fig:fields} by their cluster numbers: F170, F174, F177 and F280.
We omit the G1 field from this analysis, since at 34 kpc from the center
of M31, the frame is dominated by cluster stars.

\begin{figure}[htb]
\epsscale{1}
\plotone{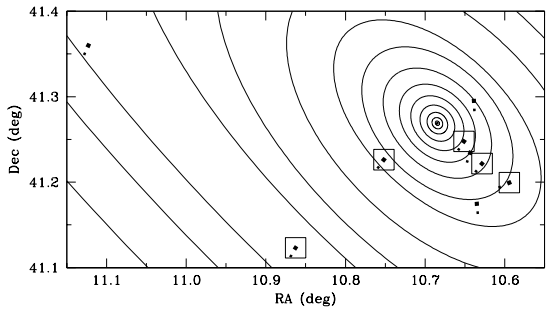} 
\figcaption{
Map of our observations.  The filled squares represent our NIC1 (small)
and NIC2 (large) fields.  The large open squares show the observations
of \citet{Rich1993}.  The contours are $r$-band surface brightness
contours taken from \citet{Kent1989} and interpolated to 0.5 magnitude
intervals.  This figure illustrates the relative positions, sizes and
separations of the NIC1 and NIC2 fields, as well as why it is not
completely fair to compare parallel NIC1 and NIC2 fields under the
assumption that they have equal surface brightness.
\label{fig:sbmap}}
\end{figure}

Since the NIC1 and NIC2 cameras are at different positions in the HST
focal plane, we can simultaneously image 2 fields at each pointing (see
Figure \ref{fig:sbmap}).  The NIC2 field is $19.2''$ across, and
separated by $17.5''$ from the $11''$ NIC1 field ($32.6''$ between field
centers).  Their different sizes are due to their different spatial
resolutions; the NIC1 images have a plate scale of $0.043''$
pixel$^{-1}$, compared to the $0.0757''$ pixel$^{-1}$ of NIC2.  Having
different resolutions at each pointing will prove useful for
understanding the severe crowding very near the center of M31.

Our observations are summarized in Table \ref{tab:observations}.  The
top half of the table lists our NIC1 observations, and the second half
is for NIC2.  The first column lists the field ID; the second and third
give the field coordinates.  The distance from the center of M31 in
arcminutes is listed in column 4.  Using this distance, the position
angle from the major axis of M31, the $r$-band surface brightness from
\citet{Kent1989}, and an assumed $(r-K)$ color of 2.9, we estimate the
$K$-band surface brightness of each field, which is listed in column 5.
Taking Kent's bulge -- disk decomposition, we also give the bulge/disk
ratio in column 6.

\begin{deluxetable}{cccccc}
\tablewidth{0pt}
\tablecaption{M31 Observations}
\tabletypesize{\footnotesize}
\tablehead{
\colhead{ID}				&
\colhead{$\alpha$\tablenotemark{a}} 	&
\colhead{$\delta$\tablenotemark{a}}	&
\colhead{Radius}			&
\colhead{$\mu_K$\tablenotemark{b}}	&
\colhead{Bulge/Disk}	\\
\colhead{(1)}		&
\colhead{(2)}		&
\colhead{(3)}		&
\colhead{(4)}		&
\colhead{(5)}		&
\colhead{(6)}		}
\startdata
NIC1 &                  &                      &          &       &     \\
F1   & 0\h 42\m 37\fs83 & 41\degr $14' 18.3''$ & $ 2.20'$ & 15.09 & 6.4 \\
F2   & 0\h 43\m 28\fs46 & 41\degr $ 6' 48.9''$ & $12.49'$ & 18.72 & 0.1 \\
F3   & 0\h 42\m 32\fs63 & 41\degr $12' 45.8''$ & $ 4.03'$ & 15.91 & 3.5 \\
F4   & 0\h 43\m  2\fs13 & 41\degr $13'  2.0''$ & $ 4.57'$ & 16.88 & 1.3 \\
F5   & 0\h 42\m 25\fs51 & 41\degr $11' 39.0''$ & $ 5.72'$ & 16.37 & 2.1 \\
F170 & 0\h 42\m 32\fs14 & 41\degr $ 9' 51.5''$ & $ 6.69'$ & 16.68 & 1.7 \\
F174 & 0\h 42\m 33\fs15 & 41\degr $17'  4.5''$ & $ 2.29'$ & 15.52 & 4.9 \\
F177 & 0\h 42\m 35\fs30 & 41\degr $13' 27.4''$ & $ 3.17'$ & 15.58 & 4.7 \\
F280 & 0\h 44\m 30\fs63 & 41\degr $21'  0.7''$ & $20.55'$ & 18.35 & 0.2 \\\hline
NIC2                    &                      &          &       &     \\
F1   & 0\h 42\m 36\fs30 & 41\degr $14' 51.7''$ & $ 1.97'$ & 14.98 & 7.7 \\
F2   & 0\h 43\m 27\fs07 & 41\degr $ 7' 23.0''$ & $11.89'$ & 18.54 & 0.1 \\
F3   & 0\h 42\m 30\fs87 & 41\degr $13' 17.7''$ & $ 3.80'$ & 15.82 & 3.9 \\
F4   & 0\h 43\m  0\fs44 & 41\degr $13' 34.3''$ & $ 3.98'$ & 16.57 & 1.9 \\
F5   & 0\h 42\m 22\fs63 & 41\degr $11' 57.8''$ & $ 5.84'$ & 16.40 & 2.0 \\
F170 & 0\h 42\m 32\fs40 & 41\degr $10' 29.0''$ & $ 6.08'$ & 16.52 & 1.9 \\
F174 & 0\h 42\m 33\fs30 & 41\degr $17' 42.0''$ & $ 2.59'$ & 15.78 & 4.1 \\
F177 & 0\h 42\m 34\fs40 & 41\degr $14'  3.6''$ & $ 2.79'$ & 15.40 & 5.5 \\
F280 & 0\h 44\m 29\fs50 & 41\degr $21' 36.0''$ & $20.49'$ & 18.33 & 0.2 \\
\enddata
\tablenotetext{a}{J2000}
\tablenotetext{b}{mag/arcsec$^{-2}$, from \citet{Kent1989} assuming $(r-K)=2.9$.}
\label{tab:observations}
\end{deluxetable}

The NICMOS focus was set at the compromise position 1-2.  This is the
best focus for simultaneous observations with cameras 1 and 2.  All of
our observations used the {\sc multiaccum} mode \citep{MacKenty1997}
because of its optimization of the detector's dynamic range and cosmic
ray rejection.

Our primary (pointed) observations were taken with NIC2.
Each field was observed through three filters: F110W (0.8--1.4
\micron), F160W (1.4--1.8 \micron), and F222M (2.15--2.30 \micron).
These filters are close to the standard ground-based $J$, $H$, \& $K$
filters.  However, to maximize the depth of our parallel NIC1
observations, we only used the F110W ($J$) filter.  Total integration
times and FWHMs for each camera and filter combination are given in
Table \ref{tab:exptimes}.  The dates of the observations are listed in
Table \ref{tab:obsdates}.

All the observations implemented a spiral dither pattern with 4
positions to compensate for imperfections in the infrared array.  For
fields F1--F5 the dither steps were $0.4''$ for the $J$ and $K$ band
images, and $1.0''$ for the $H$ band images.  Fields F170--F280 are the
same, except that we used $5.0''$ dithers in $H$.

As previously mentioned, the observations of fields F170--F280
are from another proposal targeting M31's metal-rich globular clusters
\citep{Stephens2001b}.  In these observations we exclude stars 
inside radii of $1.4''$, $0.5''$, $0.6''$, and $5.0''$ around the
clusters G170, G174, G177 and G280 respectively, to avoid cluster stars.

\begin{deluxetable}{ccccc}
\tablewidth{0pt}
\tablecaption{NICMOS Exposure Times\tablenotemark{a} \& FWHMs}
\tabletypesize{\footnotesize}
\tablehead{
\colhead{}		&
\colhead{NIC1}		&
\multicolumn{3}{c}{\underline{\hspace{1.6cm} NIC2 \hspace{1.6cm}}}\\
\colhead{Fields}	&
\colhead{F110W}		&
\colhead{F110W}         &
\colhead{F160W}         &
\colhead{F222M}		}
\startdata
F1--F5	    & 4992      & 1280      & 2048      & 1664      \\
F170--F280  & 7552      & 1920      & 3328      & 2304      \\ \hline
FWHM        & $0.099''$ & $0.125''$ & $0.148''$ & $0.185''$ \\
\enddata
\tablenotetext{a}{Exposure times in seconds.}
\label{tab:exptimes}
\end{deluxetable}

\begin{deluxetable}{cccc}
\tablewidth{0pt}
\tablecaption{Observation Dates}
\tabletypesize{\footnotesize}
\tablehead{
\colhead{Field}	&
\colhead{Date}	}
\startdata
F1	& 1998/09/20 \\
F2	& 1998/09/18 \\
F3	& 1998/09/24 \\
F4	& 1998/09/23 \\
F5	& 1998/10/13 \\
F170	& 1998/08/10 \\
F174	& 1998/08/13 \\
F177	& 1998/09/08 \\
F280	& 1998/09/13 \\
\enddata
\label{tab:obsdates}
\end{deluxetable}

\begin{figure*}
\epsscale{2}
\plotone{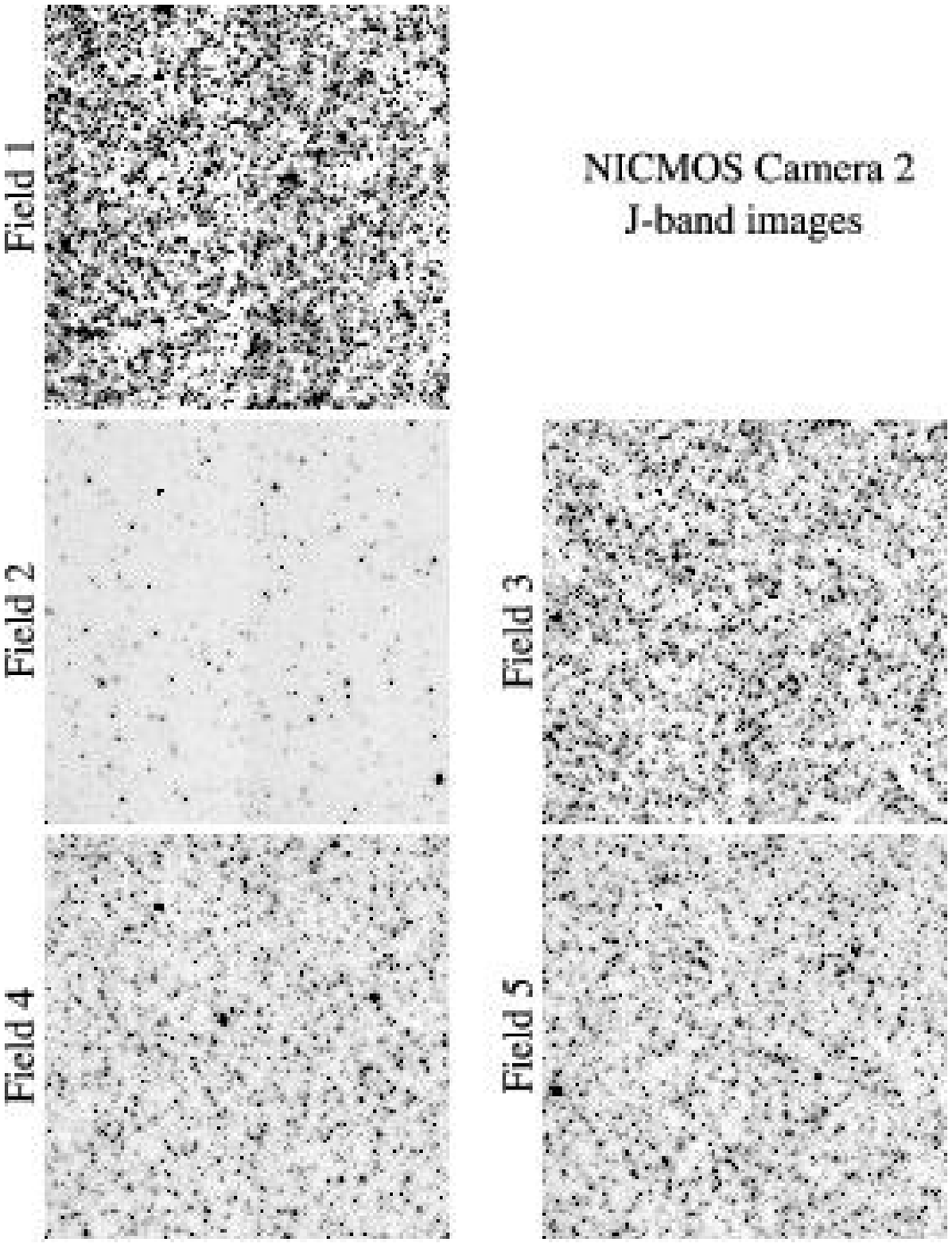} 
\figcaption{
NIC2 $J$-band images of Fields F1--F5, each image is $\sim 20''$ across.
\label{fig:7876nic2fields}}
\end{figure*}

\begin{figure*}
\epsscale{2}
\plotone{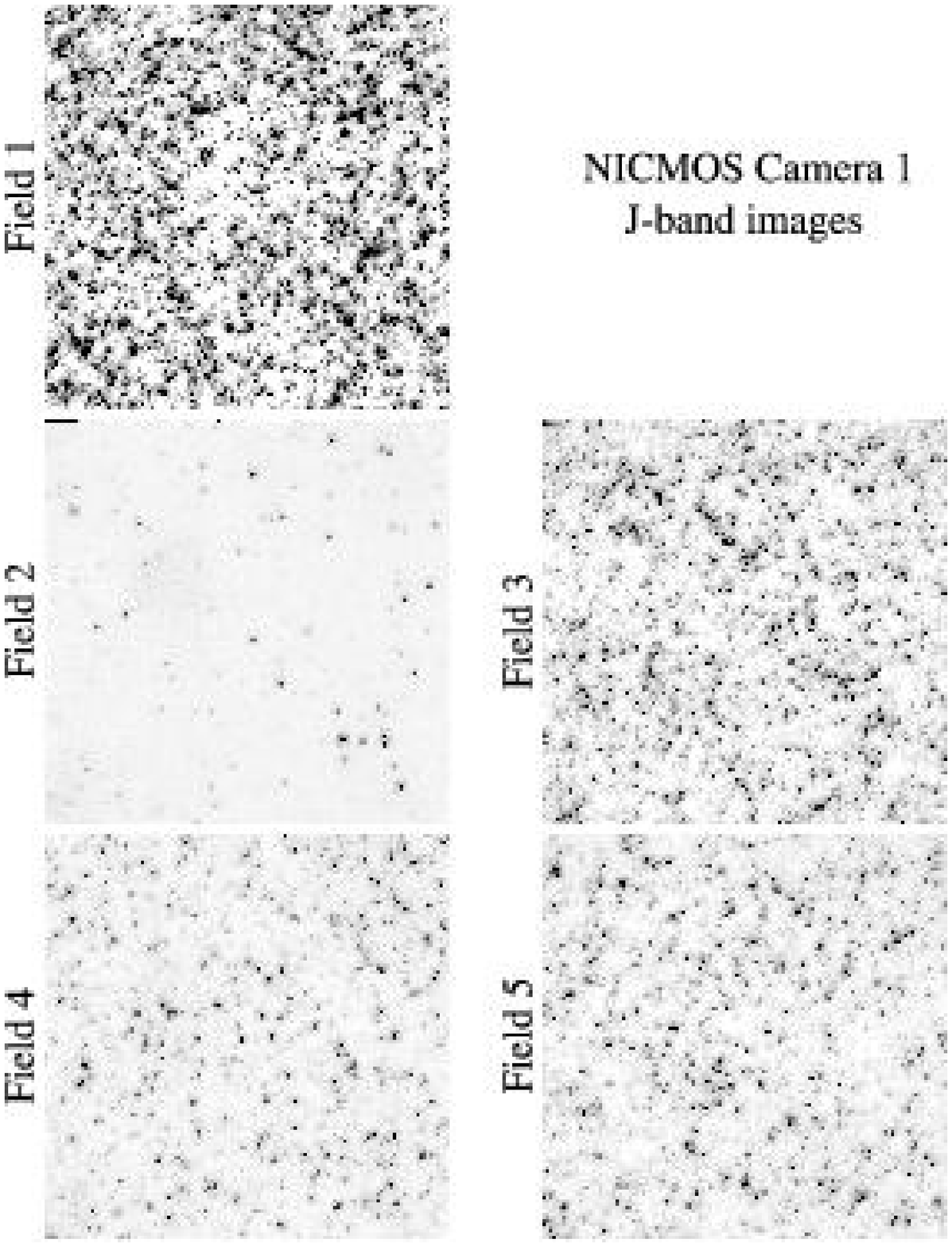} 
\figcaption{
NIC1 $J$-band images of Fields F1--F5, each image is $\sim 12''$ across.
\label{fig:7876nic1fields}}
\end{figure*}

\begin{figure*}
\epsscale{2}
\plotone{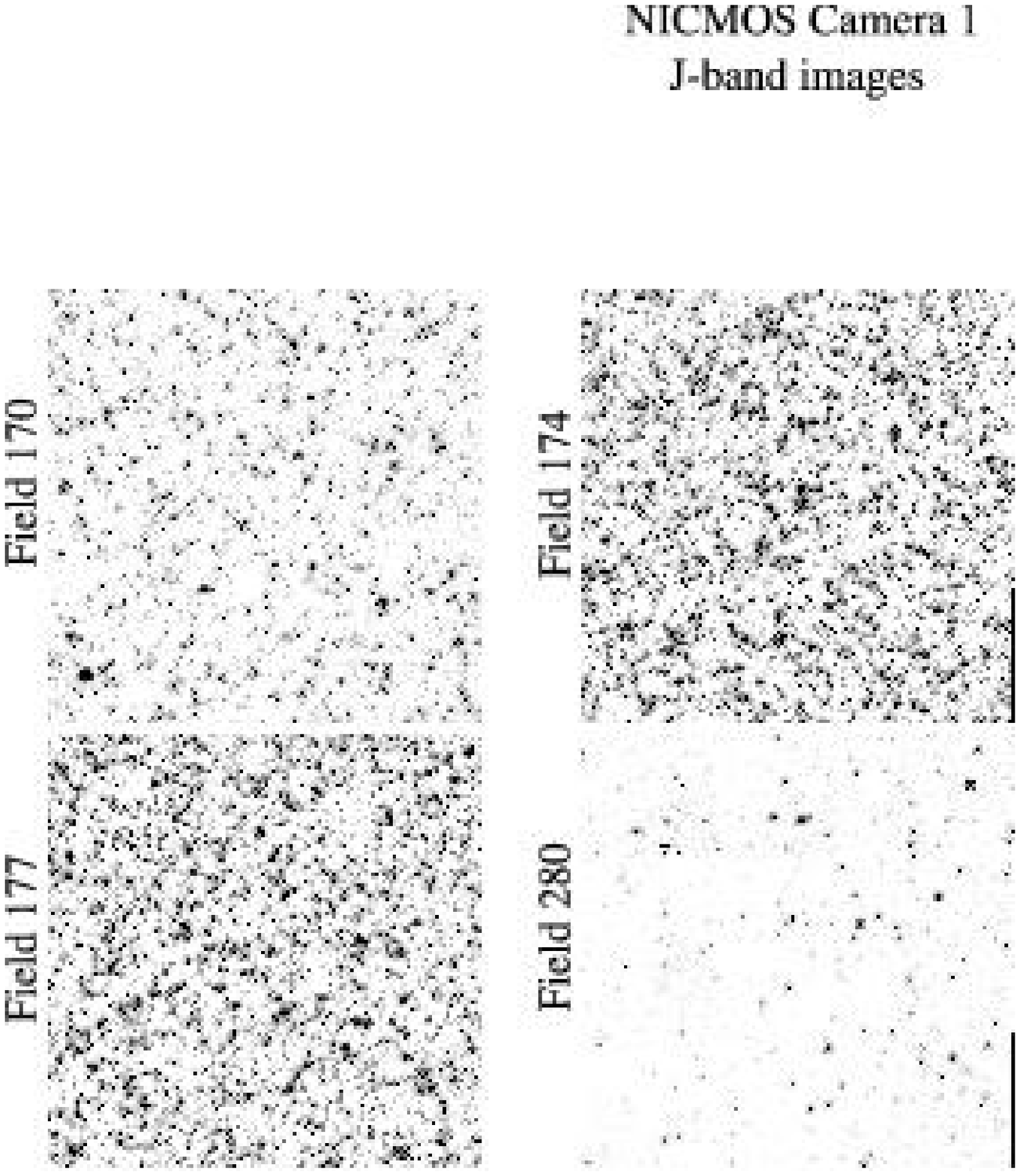} 
\figcaption{
NIC1 $J$-band images of Fields F170--F280, each image is $\sim 16''$ across.
\label{fig:7826nic1fields}}
\end{figure*}

Images of fields F1-F5 are shown in Figures \ref{fig:7876nic2fields} and
\ref{fig:7876nic1fields} for NIC2 and NIC1 respectively.  The NIC1
images of fields F170-F280 are shown in Figure \ref{fig:7826nic1fields},
and their NIC2 counterparts are given in Figure 2 of
\citet{Stephens2001b}.  These images are the combination of 4 and 12
dithers for NIC2 and NIC1 respectively.  The dimensions of each set of
combined images are different due to the varying plate scale and dither
size, and are given in the figure captions.

When converting to absolute or bolometric magnitudes we assume a
distance modulus to M31 of $(m-M)_0=24.4$, which corresponds to 3.7
parsecs arcsec$^{-1}$, and an extinction of $E(B-V)=0.22$.

\section{Data Reduction \& Photometry} \label{sec:reduction}

Our data were reduced with the STScI pipeline supplemented by the Image
Reduction and Analysis Facility (IRAF \footnote{IRAF is distributed by
the National Optical Astronomy Observatories, which are operated by
AURA, Inc., under cooperative agreement with the NSF.})  {\sc nicproto}
package (May 1999) to eliminate any residual bias (the ``pedestal''
effect).  Object detection was performed on a combined image made up of
all the dithers of all the bands (12 images in total).  Point spread
functions (PSFs) were determined from each of the four dithers, then
averaged together to create a single PSF for each band of each target.
Instrumental magnitudes were measured using the {\sc allframe} PSF
fitting software package \citep{Stetson1994}, which simultaneously fits
PSFs to all stars on all dithers.  {\sc daogrow} \citep{Stetson1990} was
then used to determine the best magnitude in a $0.5''$ radius aperture.

We finally transformed our photometry to the CIT/CTIO system.  The NIC2
measurements used the transformation equations of \citet{Stephens2000},
listed in equations 1-3.  The NIC1 transformation proved to be more
complicated, and is based on a comparison with the much lower spatial
resolution groundbased observations of \citet{Rich1993}.  A detailed
description of the technique is given in Appendix \ref{app:nic1cal}.

The selection criteria are also slightly different for stars measured in
the NIC1 and NIC2 fields.  For the NIC2 frames we require measurements
in all three bands, with PSF-fitting errors smaller than 0.25 magnitudes
in each band.  For NIC1, with only $J$-band observations, we only
require that the PSF-fitting error be less than 0.25 magnitudes.

\begin{eqnarray}
m_J = m_{110} - (0.198 \pm 0.036) (m_{110} - m_{222}) \nonumber \\ + (21.754 \pm 0.030) \\
m_H = m_{160} - (0.177 \pm 0.037) (m_{110} - m_{222}) \nonumber \\ + (21.450 \pm 0.028) \\
m_K = m_{222} + (0.074 \pm 0.037) (m_{110} - m_{222}) \nonumber \\ + (20.115 \pm 0.031)
\end{eqnarray}

\section{Color-Magnitude Diagrams} \label{sec:cmds}

\begin{figure*}[htb]
\epsscale{0.8}
\epsscale{2}
\plotone{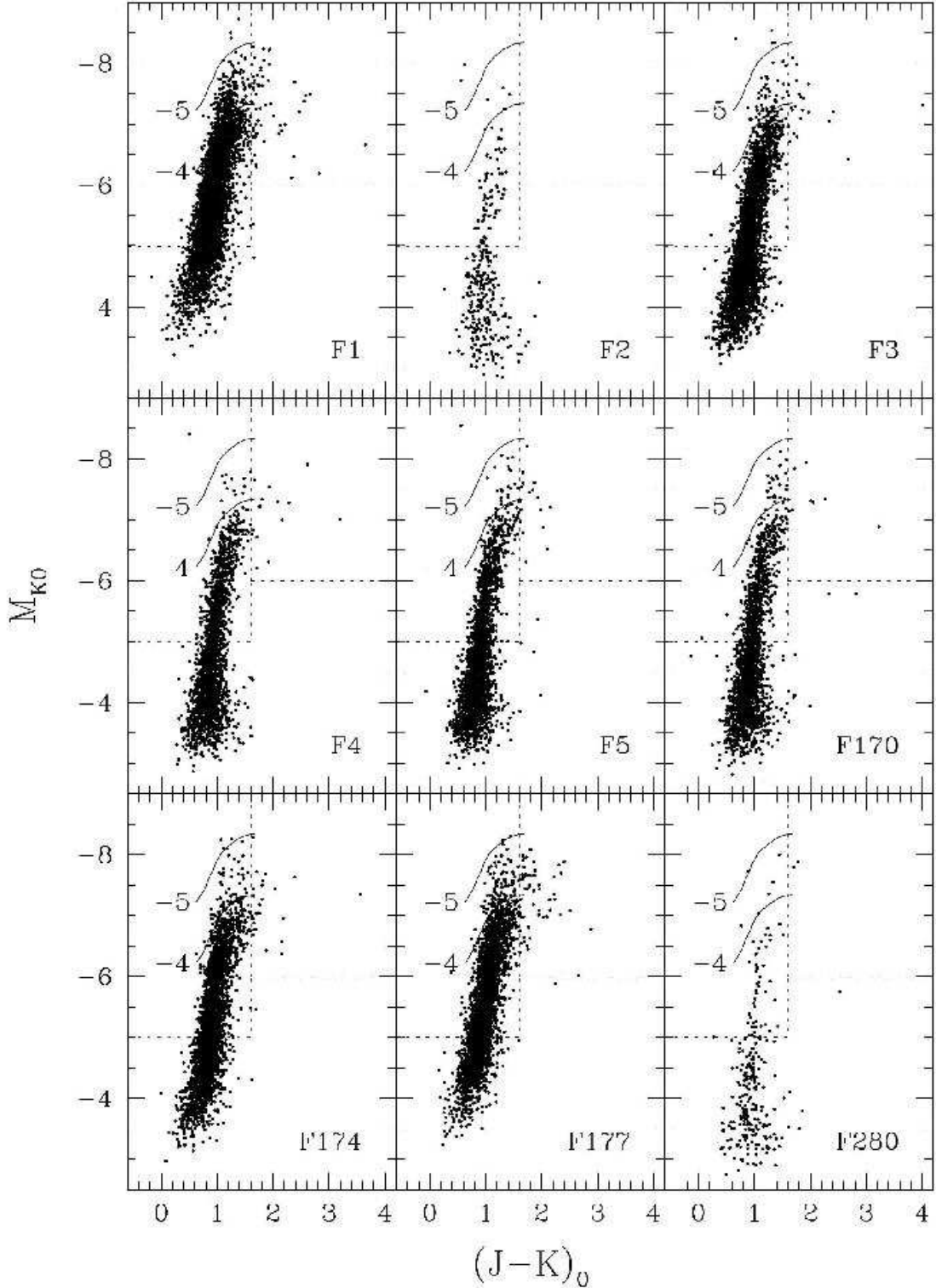} 
\figcaption{
The dereddened color magnitude diagrams for each of the 9 NIC2 fields.
We have drawn on lines of constant bolometric magnitude at $M_{bol}=-4$
and $-5$, using the bolometric corrections calculated for Baade's Window
M giants by \citet{Frogel1987}.  The box in the upper right of each
panel, with $(J-K)_0>1.6$ and $M_{K0}<-6$, indicates the region we
expect to find primarily LPVs.  The box in the upper left of each panel,
with $(J-K)_0<1.6$, and $M_{K0}<-5$, is the region we use to count
nonvariable giants.  We have assumed $E(B-V)=0.22$, giving $E(J-K)=0.12$
and $A_K=0.07$.
\label{fig:cmds}}
\end{figure*}

The $M_{K0}$--$(J-K)_0$ color-magnitude diagrams of all 9 NIC2 fields
are shown in Figure \ref{fig:cmds}.  The overplotted solid lines are
contours of constant bolometric magnitudes of $-4$ and $-5$, based on
the bolometric corrections for M giants in Baade's Window calculated by
\citet{Frogel1987}.  These plots assume a distance modulus to M31 of
$(m-M)_0=24.4$, $E(J-K)=0.12$, and $A_K=0.07$.

The RGB and AGB are both visible in these CMDs.  The tip of the RGB is
more clearly defined in the less crowded fields, and a differential
bolometric LF shows that it occurs at $M_{bol} \sim -3.75$.  In the more
crowded fields the RGB tip gets blurred because of blending, which
pushes stars up off of the RGB.  The tip of the bulge AGB extends to
$M_{bol} \sim -5$. This AGB tip is significantly fainter than previously
claimed, and we address this in a comparison with previous observations
in Section \ref{sec:comparison}.

\subsection{LPVs} \label{sec:lpvs}

Long-period variables (LPVs) are large amplitude, luminous red variable
stars with periods ranging from 50 to several hundred days.  These stars
are on the AGB and represent the brief final stages of low- to
intermediate-mass stellar evolution on the giant branch.  Based on
measurements of variables in the Galactic Bulge \citep{Frogel1987}, we
have marked the region of each CMD where we expect to find primarily
LPVs (Fig. \ref{fig:cmds}).  This region is indicated by a dashed box in
the upper right of each CMD, with $(J-K)_0>1.6$ and $M_{K0}<-6$.  Since
image blending generally shifts objects to bluer colors
\citep{Stephens2001a}, this region should be relatively insensitive to
crowding causing spurious LPV candidates, although in extreme cases,
LPVs may actually be shifted blueward {\it out} of the box, thus only
giving us a lower limit to the number of LPVs.  A casual comparison of
this LPV region between fields shows that some of the CMDs, particularly
from the inner fields, have many more potential LPVs.

Since the relative numbers and luminosities of LPVs are sensitive to the
age and/or metallicity of the parent population, LPVs can be used to
look for field-to-field variations in the stellar population. 
\citet{Frogel1998} argue that the relative number of LPVs to
non-variable giants is independent of [Fe/H] for [Fe/H]$<0$, and for
higher metallicities the LPV lifetime is significantly reduced due to
increased mass loss rates.  Thus for stellar systems with a super-solar
metallicity component, the ratio of non-variables to LPVs will appear
high compared to lower metallicity systems.  This idea seems to be
supported by their determination of a higher non-variable giants to LPV
ratio in the Galactic bulge compared to globular clusters.

\begin{figure}[htb]
\epsscale{1}
\plotone{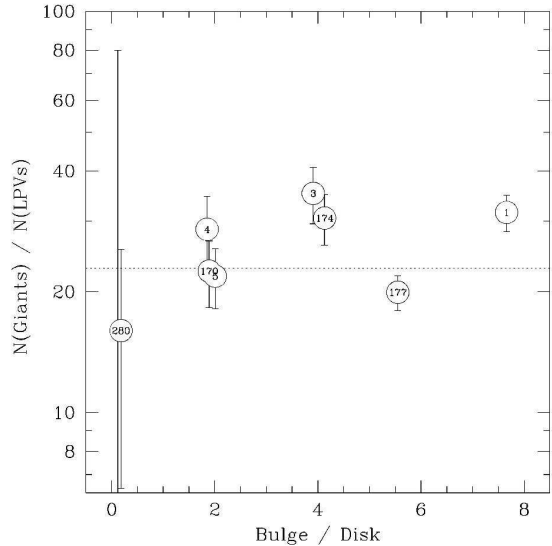} 
\figcaption{
The ratio of the number of nonvariable giants to long-period variable
(LPV) candidates as a function of bulge-to-disk ratio.  Classifications
as an LPV or non-variable giant are based on the stellar colors and
luminosities indicated on each CMD in Figure \ref{fig:cmds}.  All fields
are consistent with the average ratio of 24.  The errorbar with the
lowest bulge/disk ratio belongs to field F2, which has 80 giants, but no
LPVs.
\label{fig:lpvs}}
\end{figure}

To determine whether there is a change in the relative numbers of LPV
candidates among our fields, we compare their numbers to the number of
nonvariable giants, classified by having $(J-K)_0<1.6$ and $M_{K0}<-5$.
The ratio of giants to LPVs for each of our NIC2 fields is shown in
Figure \ref{fig:lpvs} as a function of bulge/disk ratio.  This plot
shows no trend in the giant/LPV ratio, instead all of our observations
are scattered around the average ratio of N(Giants)/N(LPVs)$\sim 24$.
Thus the LPVs appear to be uniformly distributed with the nonvariable
stars.

\clearpage		

\section{Luminosity Functions} \label{sec:lfs}

\begin{figure*}[htb]
\epsscale{2}
\plotone{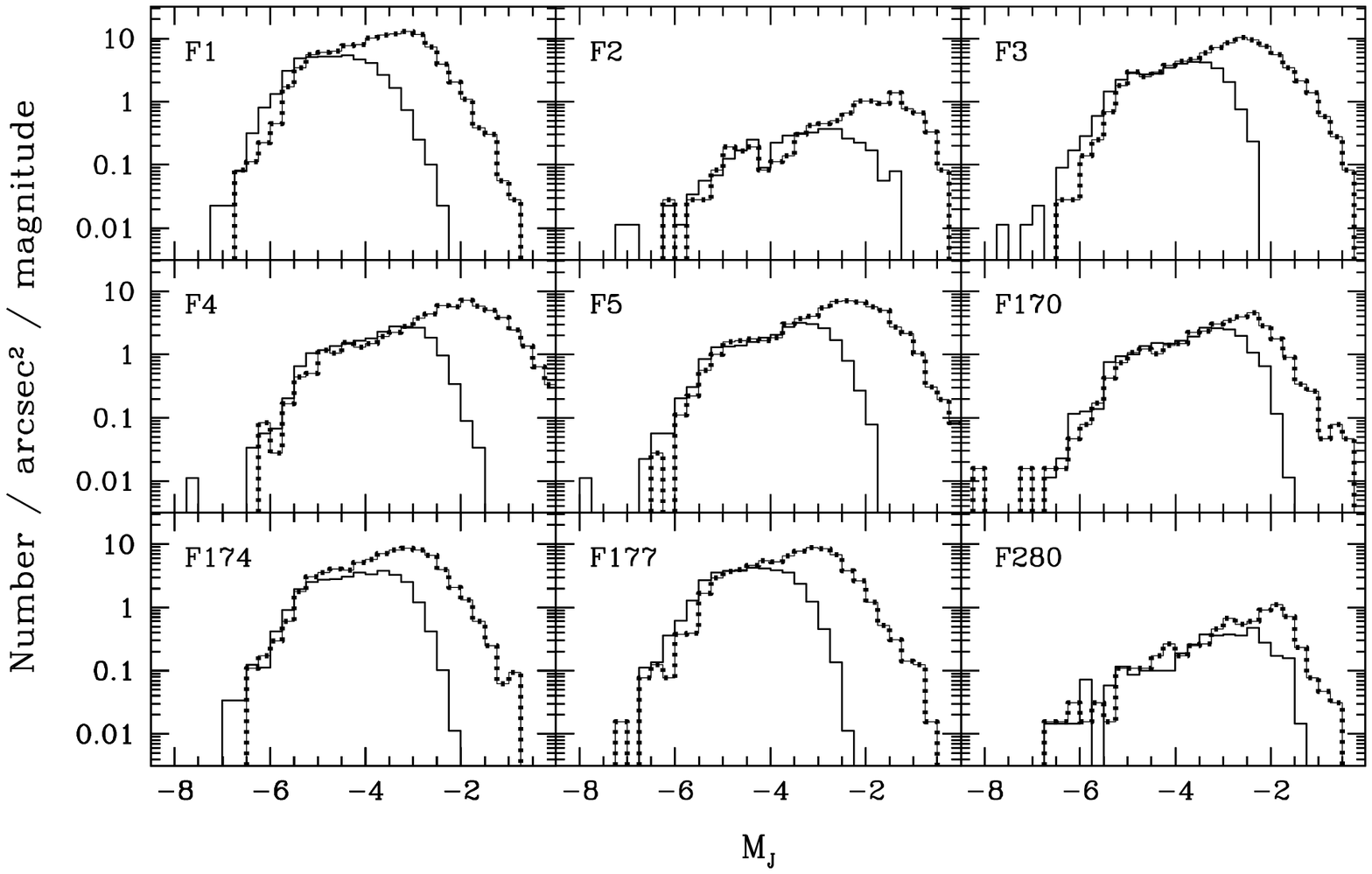} 
\figcaption{
The 18 $J$-band luminosity functions measured for our 9 sets of NIC1
(beaded) and NIC2 (solid) observations.  All of the LFs have been
normalized to show the number of stars per square arcsec per magnitude.
The NIC1 LFs go deeper because of slightly better resolution and nearly
four times longer exposures.  We have assumed a distance modulus of
$(m-M)=24.4$.
\label{fig:jlfs}}
\end{figure*}

The $J$-band LFs measured in all 18 fields are shown in Figure
\ref{fig:jlfs} and listed in Table \ref{tab:jlfs} in units of number per
magnitude per square arcsecond.  The figure shows both the NIC1 and NIC2
LFs overplotted for each field.  The first thing to note about this
compilation of M31 LFs is that the NIC1 and NIC2 measurements are not
exactly the same.  While most show good agreement, several of the NIC2
LFs extend to brighter magnitudes, and all of the NIC1 LFs extend fainter.

The faint end differences are a combined result of the longer exposure
times and better spatial resolution of NIC1.  The NIC1 exposure times
are nearly four times longer than those of the $J$-band NIC2
observations.  This is because while NIC2 was cycling through all three
$J$,$H$ \& $K$ filters, NIC1 observed only through the $J$-band filter.
The faint end photometry is also affected by the level of crowding in
the field.  NIC2 is undersampled at $J$, making it more difficult to
distinguish close objects.  Thus in very crowded fields, NIC1 has an
advantage over NIC2, accounting for the larger faint-end difference seen
in the more crowded fields.

There are several reasons for the differences seen at the bright end.
The first is the difference in the field of view of each camera.  NIC2
covers an area on the sky three times that of NIC1, and thus has a much
better chance of finding the rarer brighter stars.  Second is resolution
and image sampling.  NIC1 has a well sampled $J$-band PSF with a FWHM of
$0.099''$, while the NIC2 $J$-band PSF is undersampled with a larger
FWHM of $0.125''$.  Also, the NIC2 photometric calibration requires
detections in all three bands.  Thus the limiting resolution for NIC2 is
actually the $K$-band, which has a FWHM of $0.185''$, nearly twice the
size of the NIC1 $J$-band PSF.  The combined result is that the NIC2
observations are more sensitive to blending, which can artificially
brighten stars, extending the bright end of the LF.  Finally, there is
the coincidence that most of the NIC1 observations occur in regions of
slightly fainter surface brightnesses than their NIC2 counterparts (see
Table \ref{tab:observations} and Figure \ref{fig:sbmap}), somewhat
exacerbating both aforementioned effects.

\begin{figure*}[htb]
\epsscale{2}
\plotone{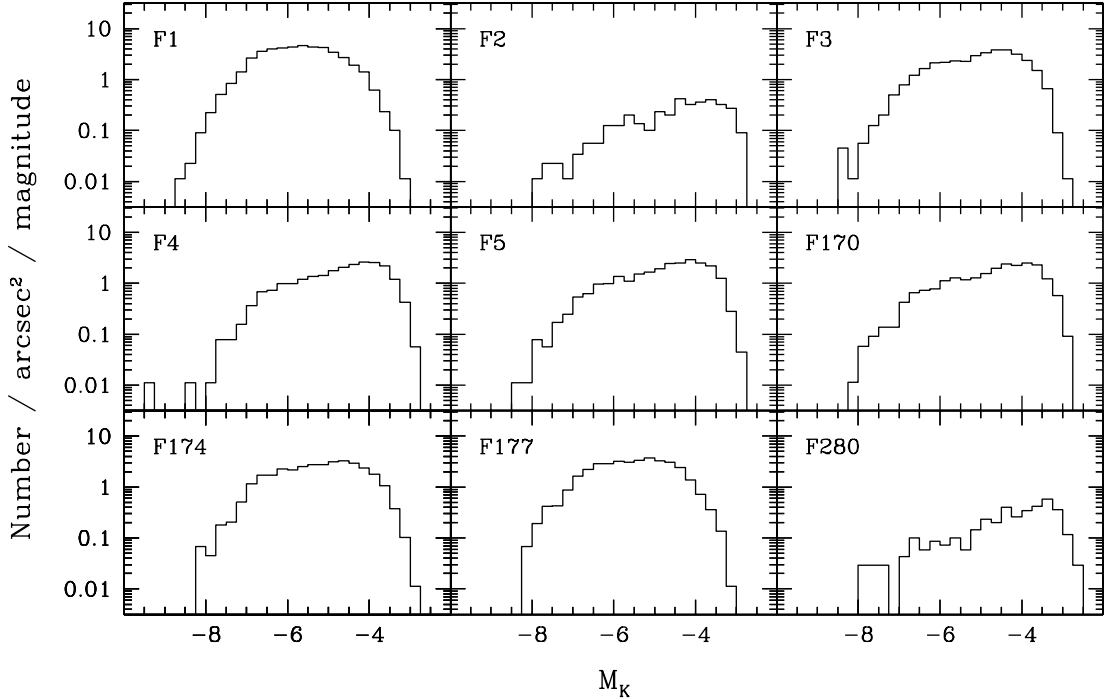} 
\figcaption{
The $K$-band luminosity functions measured with NIC2 and normalized to
give the number of stars per square arcsecond per magnitude.
\label{fig:klfs}}
\end{figure*}

The $K$-band NIC2 LFs are shown in Figure \ref{fig:klfs} and listed in
Table \ref{tab:klfs}, both normalized to give
number/arcsec$^2$/magnitude.  Although NIC1 is more resilient against
blending, both NIC1 and NIC2 observations are susceptible to its ill
effects (see \S \ref{sec:simulations}).  However, under the hypothesis
that all the measured bulge LFs arise from a single true LF, and that
the differences are purely a result of blending, we estimate that the
tip of the AGB occurs at $M_J \sim -6.25$ and $M_K \sim -8$.

\begin{deluxetable}{ccccccccccccccccccc}
\tablewidth{0pt}
\tighten
\rotate
\tablecaption{$J$-band Luminosity Functions}
\tabletypesize{\scriptsize}
\tablehead{
\colhead{}	&
\multicolumn{9}{c}{\underline{ \hspace{4cm} NIC1 \hspace{4cm} } } &
\multicolumn{9}{c}{\underline{ \hspace{4cm} NIC2 \hspace{4cm} } } \\
\colhead{$J$} & 
\colhead{F1}	&
\colhead{F2}	&
\colhead{F3}	&
\colhead{F4}	&
\colhead{F5}	&
\colhead{F170}	&
\colhead{F174}	&
\colhead{F177}	&
\colhead{F280}	&
\colhead{F1}	&
\colhead{F2}	&
\colhead{F3}	&
\colhead{F4}	&
\colhead{F5}	&
\colhead{F170}	&
\colhead{F174}	&
\colhead{F177}	&
\colhead{F280}	}
\startdata
15.875 &  0.00 &  0.00 &  0.00 &  0.00 &  0.00 &  0.00 &  0.00 &  0.00 &  0.00 &  0.00 &  0.00 &  0.00 &  0.01 &  0.00 &  0.00 &  0.00 &  0.00 &  0.00 \\ 
16.125 &  0.00 &  0.00 &  0.00 &  0.00 &  0.00 &  0.00 &  0.00 &  0.00 &  0.00 &  0.00 &  0.00 &  0.00 &  0.00 &  0.00 &  0.00 &  0.00 &  0.00 &  0.00 \\ 
16.375 &  0.00 &  0.00 &  0.00 &  0.00 &  0.00 &  0.02 &  0.00 &  0.00 &  0.00 &  0.00 &  0.00 &  0.00 &  0.00 &  0.00 &  0.00 &  0.00 &  0.00 &  0.00 \\ 
16.625 &  0.00 &  0.00 &  0.00 &  0.00 &  0.00 &  0.00 &  0.00 &  0.00 &  0.00 &  0.00 &  0.00 &  0.00 &  0.01 &  0.01 &  0.00 &  0.00 &  0.00 &  0.00 \\ 
16.875 &  0.00 &  0.00 &  0.00 &  0.00 &  0.00 &  0.00 &  0.00 &  0.00 &  0.00 &  0.00 &  0.00 &  0.01 &  0.00 &  0.00 &  0.00 &  0.00 &  0.00 &  0.00 \\ 
17.125 &  0.00 &  0.00 &  0.00 &  0.00 &  0.00 &  0.02 &  0.00 &  0.00 &  0.00 &  0.01 &  0.01 &  0.00 &  0.00 &  0.00 &  0.00 &  0.00 &  0.00 &  0.00 \\ 
17.375 &  0.00 &  0.00 &  0.00 &  0.00 &  0.00 &  0.00 &  0.00 &  0.02 &  0.00 &  0.02 &  0.01 &  0.01 &  0.00 &  0.00 &  0.00 &  0.01 &  0.00 &  0.00 \\ 
17.625 &  0.03 &  0.00 &  0.00 &  0.00 &  0.00 &  0.00 &  0.00 &  0.02 &  0.00 &  0.02 &  0.00 &  0.02 &  0.00 &  0.00 &  0.00 &  0.03 &  0.03 &  0.01 \\ 
17.875 &  0.11 &  0.00 &  0.00 &  0.00 &  0.00 &  0.02 &  0.02 &  0.06 &  0.02 &  0.16 &  0.00 &  0.02 &  0.02 &  0.05 &  0.02 &  0.09 &  0.12 &  0.00 \\ 
18.125 &  0.11 &  0.03 &  0.06 &  0.00 &  0.03 &  0.03 &  0.17 &  0.16 &  0.05 &  0.48 &  0.01 &  0.15 &  0.02 &  0.06 &  0.06 &  0.08 &  0.17 &  0.01 \\ 
18.375 &  0.19 &  0.00 &  0.06 &  0.11 &  0.03 &  0.08 &  0.20 &  0.12 &  0.02 &  1.01 &  0.01 &  0.17 &  0.07 &  0.08 &  0.11 &  0.18 &  0.52 &  0.06 \\ 
18.625 &  0.92 &  0.03 &  0.19 &  0.00 &  0.17 &  0.09 &  0.39 &  0.39 &  0.02 &  1.70 &  0.03 &  0.38 &  0.08 &  0.25 &  0.11 &  0.56 &  0.76 &  0.03 \\ 
18.875 &  2.36 &  0.00 &  0.22 &  0.36 &  0.14 &  0.22 &  0.89 &  0.70 &  0.02 &  3.95 &  0.02 &  0.92 &  0.39 &  0.50 &  0.37 &  1.34 &  2.00 &  0.01 \\ 
19.125 &  4.14 &  0.11 &  1.11 &  0.36 &  1.08 &  0.59 &  2.59 &  2.31 &  0.05 &  4.90 &  0.08 &  1.79 &  0.81 &  1.04 &  0.81 &  2.28 &  3.01 &  0.12 \\ 
19.375 &  5.81 &  0.03 &  2.33 &  0.72 &  0.94 &  0.91 &  3.14 &  3.30 &  0.14 &  5.33 &  0.11 &  2.39 &  1.04 &  1.32 &  1.01 &  2.71 &  3.45 &  0.09 \\ 
19.625 &  6.00 &  0.22 &  2.72 &  1.47 &  1.72 &  1.05 &  4.14 &  3.86 &  0.09 &  5.12 &  0.08 &  2.83 &  1.41 &  1.50 &  1.19 &  2.86 &  4.13 &  0.06 \\ 
19.875 &  7.08 &  0.14 &  2.89 &  1.22 &  1.58 &  1.27 &  3.67 &  4.16 &  0.14 &  5.31 &  0.19 &  2.67 &  1.27 &  1.42 &  1.57 &  2.76 &  3.80 &  0.13 \\ 
20.125 &  7.69 &  0.25 &  3.39 &  1.22 &  1.92 &  0.98 &  4.33 &  4.89 &  0.19 &  5.24 &  0.24 &  3.00 &  1.60 &  1.69 &  1.25 &  3.23 &  4.19 &  0.10 \\ 
20.375 &  9.33 &  0.03 &  3.61 &  1.42 &  1.50 &  1.66 &  5.52 &  5.47 &  0.23 &  4.44 &  0.18 &  3.59 &  1.68 &  1.87 &  1.66 &  3.48 &  4.05 &  0.16 \\ 
20.625 & 10.28 &  0.14 &  4.86 &  1.75 &  2.33 &  1.61 &  5.91 &  6.00 &  0.17 &  3.61 &  0.17 &  4.03 &  2.14 &  2.29 &  1.69 &  3.48 &  3.90 &  0.19 \\ 
20.875 & 11.36 &  0.19 &  4.78 &  2.06 &  3.31 &  2.33 &  7.69 &  6.67 &  0.34 &  2.25 &  0.33 &  4.35 &  2.41 &  3.10 &  2.30 &  3.78 &  3.31 &  0.27 \\ 
21.125 & 12.50 &  0.36 &  6.25 &  2.17 &  3.97 &  2.56 &  8.77 &  8.34 &  0.33 &  1.19 &  0.34 &  3.97 &  2.68 &  3.12 &  2.65 &  2.90 &  1.76 &  0.41 \\ 
21.375 & 12.53 &  0.44 &  7.61 &  3.17 &  4.58 &  3.33 &  8.50 &  8.62 &  0.48 &  0.51 &  0.37 &  2.74 &  2.75 &  2.79 &  2.66 &  2.21 &  0.88 &  0.30 \\ 
21.625 &  9.58 &  0.50 &  9.25 &  4.06 &  5.94 &  3.58 &  7.88 &  8.36 &  0.67 &  0.17 &  0.30 &  1.60 &  2.35 &  2.29 &  2.39 &  0.70 &  0.27 &  0.45 \\ 
21.875 &  6.36 &  0.42 & 10.56 &  5.03 &  6.83 &  4.59 &  5.41 &  5.19 &  0.56 &  0.06 &  0.35 &  0.48 &  1.58 &  1.44 &  1.69 &  0.25 &  0.05 &  0.41 \\ 
22.125 &  3.00 &  0.94 &  8.89 &  6.08 &  7.47 &  3.69 &  3.12 &  3.48 &  0.64 &  0.00 &  0.28 &  0.11 &  0.68 &  0.44 &  0.78 &  0.06 &  0.01 &  0.41 \\ 
22.375 &  1.69 &  1.00 &  6.69 &  6.14 &  6.28 &  2.44 &  1.69 &  1.97 &  1.12 &  0.00 &  0.19 &  0.00 &  0.19 &  0.16 &  0.34 &  0.01 &  0.00 &  0.20 \\ 
22.625 &  0.47 &  1.03 &  4.36 &  6.72 &  5.36 &  1.28 &  0.89 &  1.09 &  0.97 &  0.00 &  0.14 &  0.00 &  0.08 &  0.03 &  0.09 &  0.00 &  0.00 &  0.22 \\ 
22.875 &  0.36 &  1.06 &  2.69 &  5.78 &  4.28 &  0.73 &  0.56 &  0.36 &  0.48 &  0.00 &  0.07 &  0.00 &  0.00 &  0.00 &  0.00 &  0.00 &  0.00 &  0.06 \\ 
23.125 &  0.19 &  1.03 &  1.86 &  4.44 &  2.28 &  0.36 &  0.12 &  0.20 &  0.14 &  0.00 &  0.02 &  0.00 &  0.00 &  0.00 &  0.00 &  0.00 &  0.00 &  0.01 \\ 
23.375 &  0.03 &  0.81 &  1.11 &  3.42 &  1.81 &  0.09 &  0.11 &  0.11 &  0.09 &  0.00 &  0.00 &  0.00 &  0.00 &  0.00 &  0.00 &  0.00 &  0.00 &  0.00 \\ 
23.625 &  0.00 &  0.58 &  0.50 &  1.92 &  0.56 &  0.06 &  0.03 &  0.09 &  0.02 &  0.00 &  0.00 &  0.00 &  0.00 &  0.00 &  0.00 &  0.00 &  0.00 &  0.00 \\ 
23.875 &  0.00 &  0.28 &  0.08 &  1.00 &  0.33 &  0.06 &  0.00 &  0.00 &  0.02 &  0.00 &  0.00 &  0.00 &  0.00 &  0.00 &  0.00 &  0.00 &  0.00 &  0.00 \\ 
24.125 &  0.00 &  0.03 &  0.08 &  0.53 &  0.06 &  0.02 &  0.00 &  0.00 &  0.00 &  0.00 &  0.00 &  0.00 &  0.00 &  0.00 &  0.00 &  0.00 &  0.00 &  0.00 \\ 
\enddata
\label{tab:jlfs}
\end{deluxetable}

\begin{deluxetable}{cccccccccc}
\tablewidth{0pt}
\tablecaption{$K$-band Luminosity Functions}
\tabletypesize{\footnotesize}
\tablehead{
\colhead{$K$} & 
\colhead{F1}	&
\colhead{F2}	&
\colhead{F3}	&
\colhead{F4}	&
\colhead{F5}	&
\colhead{F170}	&
\colhead{F174}	&
\colhead{F177}	&
\colhead{F280}	}
\startdata
14.875 & 0.00 & 0.00 & 0.00 & 0.01 & 0.00 & 0.00 & 0.00 & 0.00 & 0.00 \\ 
15.125 & 0.00 & 0.00 & 0.00 & 0.00 & 0.00 & 0.00 & 0.00 & 0.00 & 0.00 \\ 
15.375 & 0.00 & 0.00 & 0.00 & 0.00 & 0.00 & 0.00 & 0.00 & 0.00 & 0.00 \\ 
15.625 & 0.01 & 0.00 & 0.00 & 0.00 & 0.00 & 0.00 & 0.00 & 0.00 & 0.00 \\ 
15.875 & 0.01 & 0.00 & 0.01 & 0.00 & 0.01 & 0.00 & 0.00 & 0.00 & 0.00 \\ 
16.125 & 0.05 & 0.00 & 0.03 & 0.01 & 0.00 & 0.00 & 0.05 & 0.02 & 0.00 \\ 
16.375 & 0.15 & 0.01 & 0.06 & 0.00 & 0.05 & 0.03 & 0.05 & 0.09 & 0.01 \\ 
16.625 & 0.32 & 0.01 & 0.07 & 0.05 & 0.06 & 0.06 & 0.07 & 0.26 & 0.04 \\ 
16.875 & 0.61 & 0.02 & 0.17 & 0.09 & 0.15 & 0.11 & 0.18 & 0.50 & 0.01 \\ 
17.125 & 1.06 & 0.02 & 0.32 & 0.08 & 0.17 & 0.14 & 0.34 & 0.53 & 0.01 \\ 
17.375 & 1.71 & 0.00 & 0.52 & 0.21 & 0.36 & 0.19 & 0.64 & 1.25 & 0.01 \\ 
17.625 & 3.18 & 0.06 & 0.96 & 0.50 & 0.59 & 0.60 & 1.46 & 1.79 & 0.04 \\ 
17.875 & 3.72 & 0.07 & 1.41 & 0.69 & 0.78 & 0.73 & 1.69 & 2.48 & 0.12 \\ 
18.125 & 4.24 & 0.03 & 1.81 & 0.83 & 0.88 & 0.64 & 1.94 & 2.81 & 0.04 \\ 
18.375 & 4.15 & 0.15 & 2.16 & 0.98 & 1.19 & 0.88 & 2.30 & 3.13 & 0.09 \\ 
18.625 & 4.36 & 0.18 & 2.40 & 1.15 & 1.19 & 1.38 & 2.22 & 3.18 & 0.13 \\ 
18.875 & 4.61 & 0.19 & 2.22 & 1.16 & 1.26 & 1.12 & 2.71 & 3.18 & 0.06 \\ 
19.125 & 4.42 & 0.10 & 2.50 & 1.41 & 1.67 & 1.18 & 2.81 & 3.66 & 0.07 \\ 
19.375 & 4.01 & 0.14 & 3.10 & 1.50 & 1.69 & 1.43 & 2.72 & 3.49 & 0.23 \\ 
19.625 & 3.11 & 0.25 & 3.61 & 2.04 & 2.13 & 1.71 & 3.32 & 3.22 & 0.16 \\ 
19.875 & 2.33 & 0.28 & 3.77 & 1.95 & 2.52 & 2.08 & 3.02 & 2.91 & 0.32 \\ 
20.125 & 1.71 & 0.35 & 3.67 & 2.61 & 2.93 & 2.49 & 2.74 & 1.98 & 0.36 \\ 
20.375 & 1.10 & 0.38 & 2.84 & 2.55 & 2.52 & 2.28 & 2.19 & 1.01 & 0.27 \\ 
20.625 & 0.41 & 0.37 & 2.16 & 2.47 & 2.41 & 2.53 & 1.55 & 0.60 & 0.36 \\ 
20.875 & 0.19 & 0.39 & 1.05 & 1.79 & 2.05 & 1.85 & 0.75 & 0.26 & 0.58 \\ 
21.125 & 0.03 & 0.33 & 0.37 & 1.01 & 0.68 & 1.00 & 0.21 & 0.07 & 0.38 \\ 
21.375 & 0.01 & 0.17 & 0.03 & 0.17 & 0.14 & 0.30 & 0.05 & 0.00 & 0.29 \\ 
\enddata
\label{tab:klfs}
\end{deluxetable}

\subsection{Comparison with Baade's Window} \label{sec:compbw}

We show our NIC2 $K$-band luminosity functions superposed on the LF of
the Galactic Bulge in Figure \ref{fig:compbw}.  The Bulge LF is a
composite of measurements made in Baade's Window (BW), with the bright
end ($M_K < -6.6$) from \citet{Frogel1987} and the faint end from
\citet{DePoy1993}.  All the LFs have been normalized in the range $-7 <
M_K < -5.5$.

The bottom panel of Figure \ref{fig:compbw} is a comparison between BW
and the seven bulge fields in our sample.  These fields all have
bulge-to-disk ratios greater than one, as listed in Table
\ref{tab:observations}.  These fields are all of very high surface
brightness, and therefore we expect most to exhibit some amount of
artificial brightening due to blending.  However, as the plot shows,
there is still very good agreement between these M31 bulge fields and
the bulge of the MW.

Thus based on our infrared luminosity functions, the stellar population
of bulge of M31 is very similar to that of the Milky Way.  The match up
between the LFs is quite good, and when one takes into account our
prediction of a small amount of artificial brightening due to blending
in our most crowded bulge fields, the correspondence will be even better.

\subsection{Comparison of Disk and Bulge LFs} \label{sec:compdb}

\begin{figure}[htb]
\epsscale{1}
\plotone{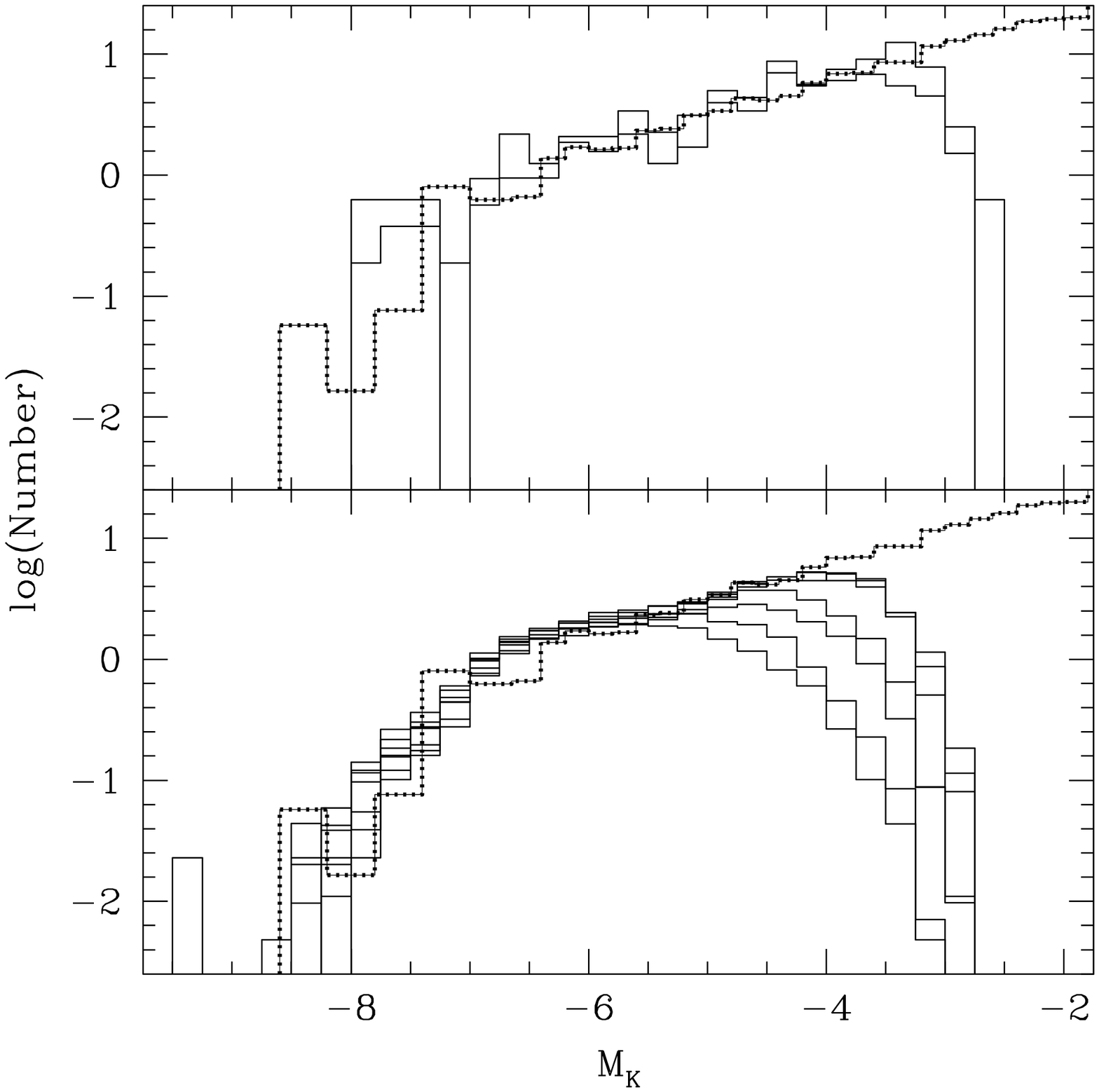} 
\figcaption{
Comparison of M31 luminosity functions with that of Baade's Window (BW).
The top panel shows the M31 disk (F2 \& F280) LFs (solid) compared with
the BW LF (beaded).  The bottom panel shows the M31 bulge LFs (solid)
overplotted on the BW LF (beaded).  All LFs have been normalized in the
range $-7 < M_K < -5.5$.  The BW LF is a composite of measurements made
by \citet{Frogel1987} and \citet{DePoy1993}.  Note that the M31 disk LFs
extend $>0.5$ magnitudes brighter than the cutoff at $M_K \sim -7.4$
seen in BW, while the M31 bulge LFs are in good agreement with the
observations of BW.
\label{fig:compbw}}
\end{figure}

The top panel of Figure \ref{fig:compbw} is a comparison between BW and
the two disk fields in our sample.  The F2 \& F280 fields both have
bulge-to-disk ratios less than one (see Table \ref{tab:observations}).
These two fields also have the lowest surface brightnesses, which means
that the effects of blending are the least, and hence their photometric
measurements are the most trustworthy.  However in this comparison, we
see that both of these disk LFs extend slightly brighter than the break
measured in BW, and do so more prominently than any of the bulge fields.

In order to determine whether or not the measured disk and bulge
luminosity functions are in fact distinguishable from one another, we
compare the distributions of stellar luminosities using the KS-test.  We
combine the F2 and F280 measurements to represent the disk population,
and use all other fields for the bulge population.  Incompleteness in
the more crowded fields limits the comparison to only bright stars with
$M_K < -6$, and in this range the KS-test shows a conspicuous
overabundance of luminous disk stars in the range $-8 < M_K < -7.5$.
However the significance is low, with $P=0.34$, indicating that the two
populations are nonetheless consistent with being drawn from the same
parent population.

On the other hand, if we limit the comparison to only the AGB ($M_K <
-7$), the excess of luminous disk stars is enough to drop the KS
$P-$value to $0.02$.  This low probability is marginally significant,
but is based on a much smaller number of stars (12 disk stars and 729
bulge stars).  It is also noteworthy that the simulations (\S
\ref{sec:simulations}) show no such enhancement.

In summary, both of the two disk fields have a slight excess of luminous
stars with $-8 < M_K < -7.5$, although only statistically significant
when compared to the bulge fields over a small range in luminosity.
This small overabundance of AGB stars just above the Baade's Window LF
break is due to the presence of younger disk stars in these two fields.

\section{Blending} \label{sec:blending}

In order to analytically estimate the effects of blending on our
observations, we have used the equations of \citet{Renzini1998} to
predict number of stars in each evolutionary stage per resolution
element.  Several parameters in this calculation have a weak dependence
on the age, metallicity, and IMF of the assumed stellar population.  For
these parameters, we choose to use a ratio of total to $K$-band
luminosity of $L_T / L_K = 0.36$, and a specific evolutionary flux
$B(t)= 2.2 \times 10^{-11}$ stars yr$^{-1}$ \lsun$^{-1}$, both suitable
for a solar-metallicity, 15 Gyr old population.

To try to estimate the importance of blending on fields at different
distances from M31, we have calculated the number of RGB stars within 1
magnitude of the RGB tip, $N(RGBT)$.  Since the brightest stars in our
fields are only $\sim 1$ magnitude brighter than the expected tip of the
RGB, a blend with even one RGBT star will distort their measurement by
$>16\%$.

\begin{figure}[htb]
\epsscale{1}
\plotone{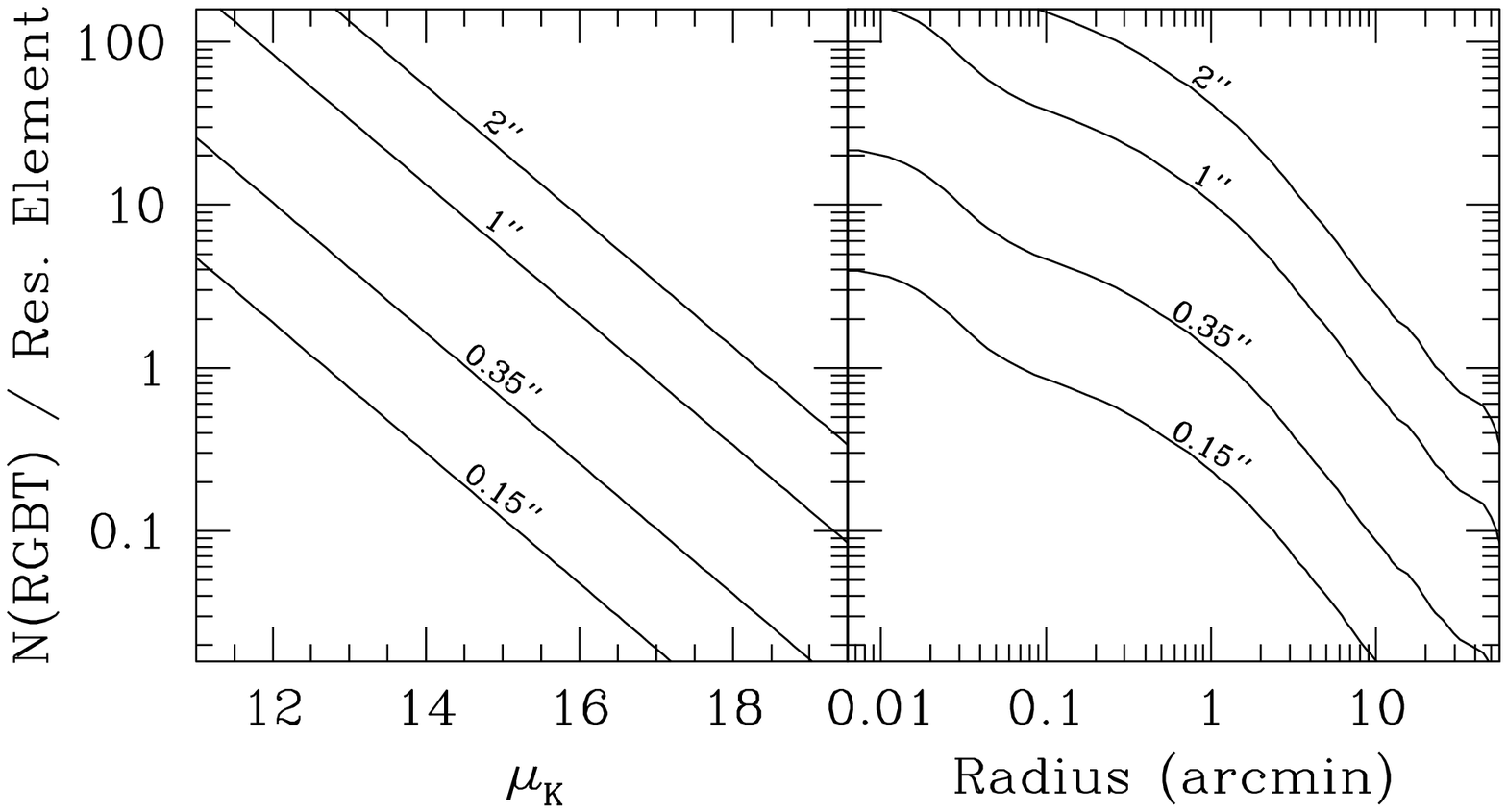} 
\figcaption{
The number of M31 RGB stars within one magnitude of the RGB tip per
resolution element, based on the formulae of \citet{Renzini1998}.  The
left panel shows N(RGBT) per resolution element as a function of the
$K$-band surface brightness (magnitudes arcsec$^{-2}$) for four
different imaging resolutions.  The right panel shows N(RGBT) per
resolution element as a function of the distance from the center of M31
in arcminutes, based on the major-axis surface brightness measurements
of \citet{Kent1989}.  The transformation from Kent's $r$-band
measurements to $K$-band surface brightness assume a constant color of
$(r-K)=2.9$.
\label{fig:nrgbt_reselement}}
\end{figure}

The results of these calculations are displayed in Figure
\ref{fig:nrgbt_reselement}.  The left panel shows the number of RGBT
stars per resolution element as a function of surface brightness for
four different imaging resolutions.  The $0.15''$ resolution roughly
corresponds to NICMOS, $0.35''$ is the resolution of the
\citet{Davidge2001} observations, and $1''$ corresponds to the
resolution obtained by \citet{Rich1993}.  As an example of using this
plot, consider an image with $0.35''$ resolution taken at a location
where the $K$-band surface brightness is 14.5 magnitudes
arcsecond$^{-1}$.  That image would have, on average, 1 RGBT star in
each resolution element; obviously not conditions favorable for accurate
photometry.

To make this plot easier to interpret, on the right side of Figure
\ref{fig:nrgbt_reselement} we show the number of RGBT stars per
resolution element as a function of position in M31, using the same four
imaging resolutions.  To convert from the $K$-band surface brightness in
the left panel to radius in the right panel, we have used the $r$-band
surface brightness measurements of \citet{Kent1989}, and we assume that
$(r-K)=2.9$.  Following through with the previous example, we see that
with $0.35''$ resolution, we would find approximately 1 per resolution
element at a distance of $\sim 1.3'$ from the center of M31 along its
major axis.

The question then becomes, what is the limit for ``good'' photometry?
While it seems quite obvious what would be bad, i.e. one or more RGBT
star per resolution element, what is ``good'' is more difficult to
quantify, and requires knowledge of exactly how good ``good'' needs to
be.  Clearly, if $N(RGBT)$ is less than one, $N(RGBT)$ is approximately
the probability that a star within one magnitude of the RGB tip will
fall in any given resolution element.  Thus $N(RGBT)^2$ is the
probability that a resolution element will contain a blend of two RGBT
stars.  Therefore the number, and severity of the blends which can be
accepted determine the limiting surface brightness.
\citet{Stephens2001a} have run simulations on their NICMOS photometry of
globular clusters in M31, and find that for accurate photometry of stars
down to $K \sim 21$, $N(RGBT)$/resolution element should be $\lesssim
0.05$.  While this is a good guide, to better understand the
observations at different resolutions and surface brightnesses, it is
best to perform simulations to attempt to quantify the effects of
blending.

\section{Simulations} \label{sec:simulations}

To better understand the effects of blending we have run extensive
simulations of each of our NICMOS fields following the procedures of
\citet{Stephens2001a}.  We create an artificial field to match each
observed field, and measure it in exactly the same manner as the real
frame.  Since in the simulations we know both the measured and true
magnitude of every star in the field, we can try to estimate the true
properties of the observed stellar population being modeled, free from
observational effects.

One of the goals of this work was to look for variations in the stellar
populations with varying galactocentric distance and bulge-to-disk
ratio.  However, the severe crowding, strong dependence of blending on
surface brightness, and degeneracy between surface brightness and
bulge-to-disk ratio make this question very difficult to answer.  Thus
one of the main purposes of our simulations is to determine whether all
of our observations are consistent with a single stellar population.  To
make this determination, we have generated artificial frames using the
stellar properties measured in some of the least crowded fields.  If the
measured differences between fields are just due to observational
effects, the simulations should exhibit the same differences.

The simulations are complicated by the fact that the data come from four
different instrument configurations, each with different exposure times,
plate scales, and dither sizes.  For {\it each} configuration, each
dither starts as a blank frame having the appropriate noise
characteristics.  We then randomly add stars using the {\sc daophot}
{\sc addstar} routine, until we have approximately matched the observed
stellar density in the field being modeled.  The PSFs used to add stars
are the average of the PSFs determined from each field for each
configuration, with any negative values in the model PSF set to zero.
The {\sc addstar} routine also incorporates random Poisson noise into
each star as it is added.

The input stellar population was chosen to match the colors and
luminosity function observed in the least-crowded bulge fields.  The
colors are the mean colors observed in fields 4 \& 5, calculated at 0.5
magnitude intervals.  The input LF is a broken power law with a faint
end slope of 0.278 extending from $-5.7< M_K <5.0$, and a bright-end
slope of 1.100 from $-7.4< M_K <-5.7$.  The faint-end slope was taken
from the Galactic Bulge \citep{DePoy1993}, while the breakpoint and
bright-end slope were determined from the NIC1 $J$-band LFs, since it
was felt that they were the most robust against the effects of blending.
The artificial frames were then processed and measured in exactly the
same manner as the real data, namely finding stars on a combined image
with {\sc daofind}, then measuring all dithers simultaneously with {\sc
allframe}.

The number of input stars was varied to approximately match the number
of detected stars on each frame, although the measured LF morphology was
also taken into account for some of higher density fields.  Table
\ref{tab:simulations} lists the number of artificial stars input
into each simulated frame, as well as the number of stars recovered from
both the real and simulated fields.

\begin{figure*}[htb]
\epsscale{2}
\plotone{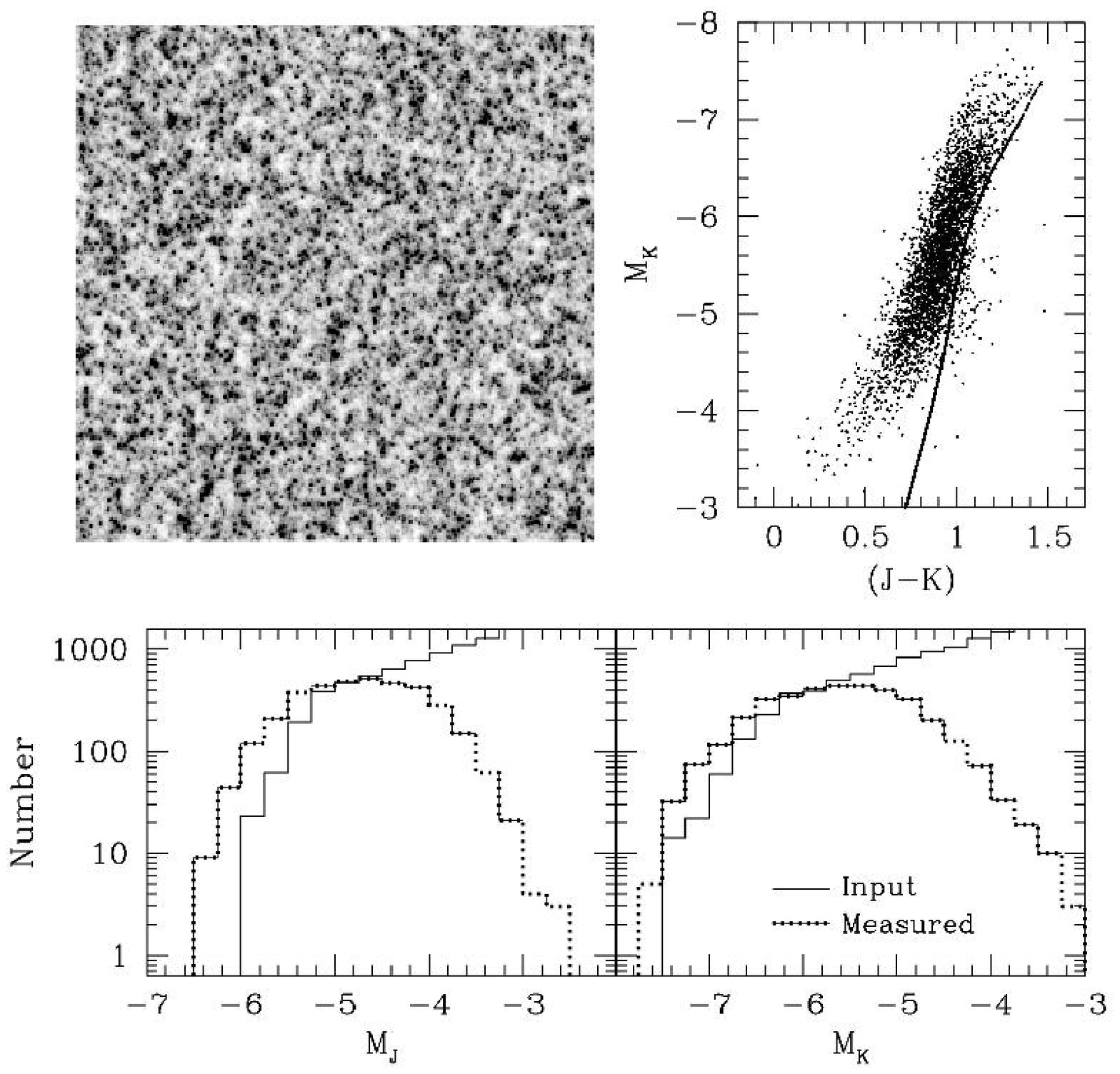} 
\figcaption{
Simulation of the NIC2 field F1.  The combined $J$-band image is
illustrated in the upper left.  The CMD showing both input (narrow locus
of stars) and measured stars is in the upper right.  The $J$ and
$K$-band input (solid) and measured (beaded) luminosity functions are
shown in the bottom two panels.
\label{fig:sims}}
\end{figure*}

The results of the simulations of a single camera (NIC2) and a single
field (F1) are summarized in Figure \ref{fig:sims}.  This is the most
crowded field, and hence the effects of blending are the most severe.
The {\em simulated} $J$-band image is shown in the upper left, and is
nearly indistinguishable from the real observations
(Fig. \ref{fig:7876nic2fields}).  The resulting $M_K$,$(J-K)$ CMD is
shown in the upper right.  The input stars form a narrow locus which
blends into a line stretching (on this plot) from $M_K=-3$ and
$(J-K)=0.7$ to $M_K=-7.4$ and $(J-K)=1.48$.  The locus of measured stars
is broader and shifted brighter and bluer compared to the input stars.
As the simulations show, when the field is this crowded, {\em none} of
our measurements are very accurate.

The input and recovered luminosity functions are shown at the bottom of
Figure \ref{fig:sims}.  The $J$-band LFs on the left show that the
measured LF is shifted by $\sim 0.5$ magnitudes brighter than the input
LF.  The LFs on the right show that, the $K$-band measurements are not
as severely distorted as the undersampled NIC2 $J$-band observations.

\subsection{Simulation Results} \label{sec:simresults}

\begin{figure}[htb]
\epsscale{1}
\plotone{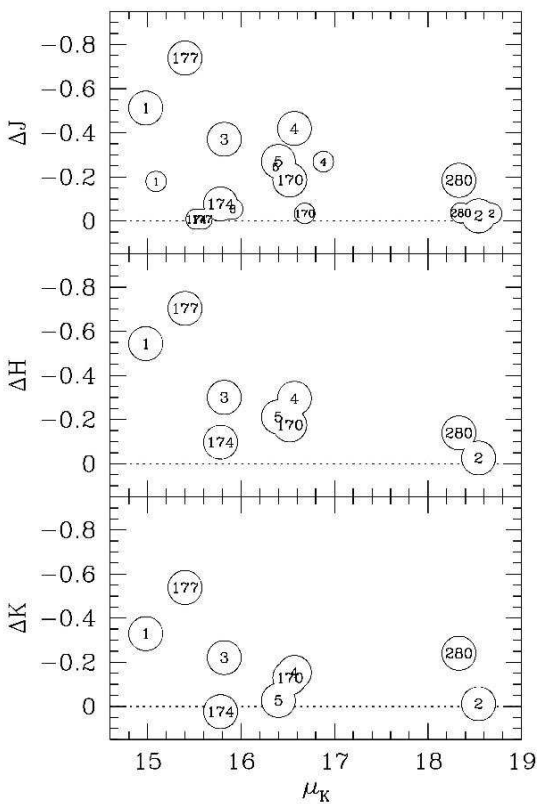} 
\figcaption{
The difference between the brightest measured and brightest input star
in each simulated field plotted as a function of the field $K$-band
surface brightness (mag/arcsec$^2$).  The top panel shows the $J$-band
difference for the NIC1 and NIC2 fields, the bottom two panels show the
$H$- and $K$-band difference for the NIC2 fields.  The Field ID is
indicated in the middle of each circle, where the large circles denote
the larger NIC2 fields, and the smaller circles the smaller NIC1 fields ($J$-band).
\label{fig:deltamag}}
\end{figure}

The first statistic we calculate is the difference between the brightest
star measured and the brightest star input into each artificial field.
This gives a rough idea of the maximum amount of brightening one can
expect in each field.  Figure \ref{fig:deltamag} illustrates this
difference for the $J$,$H$, and $K$-bands as a function of field surface
brightness.  This plot shows that there is little brightening due to
blending in the artificial fields with lower surface brightnesses
($\mu_K > 18$).  However, as the surface brightness increases, the
amount of artificial brightening increases as well.

The size of each circle in Figure \ref{fig:deltamag} indicates whether
the simulation is a NIC1 (small circles) or NIC2 (large circles) field.
The higher resolution and more finely sampled NIC1 observations are
clearly less affected by blending, with only a few fields having
deviations greater than 0.1 magnitudes.

This figure also shows that the brightest $K$-band data are less
affected by blending than the corresponding $J$-band measurements.  For
the most crowded fields (e.g. F1, F177, F174, F3) the $J$-band
brightening due to blending can be as high as 0.75 magnitudes, while in
the $K$-band, blending is $\sim 0.2$ magnitudes less.  This can also be
seen in the larger difference between the input and recovered LFs at $J$
compared to $K$, and the blueward shift of stars in the CMD as seen in
Figure \ref{fig:sims}.  As mentioned above, we attribute the difference
between $J$ \& $K$ to the undersampling in $J$ and the bluer color of
the underlying population.

The delta-magnitude quantity previously calculated is admittedly subject
to small number statistics, since it is based on only a single star in
each field.  Ideally we would run each simulation a number of times, and
the average difference would be a much more robust estimator of the
amount of brightening to expect in the real field.  However, looking at
the results of all the fields together, the direction and magnitude of
the effect is clear.

\begin{figure}[htb]
\epsscale{1}
\plotone{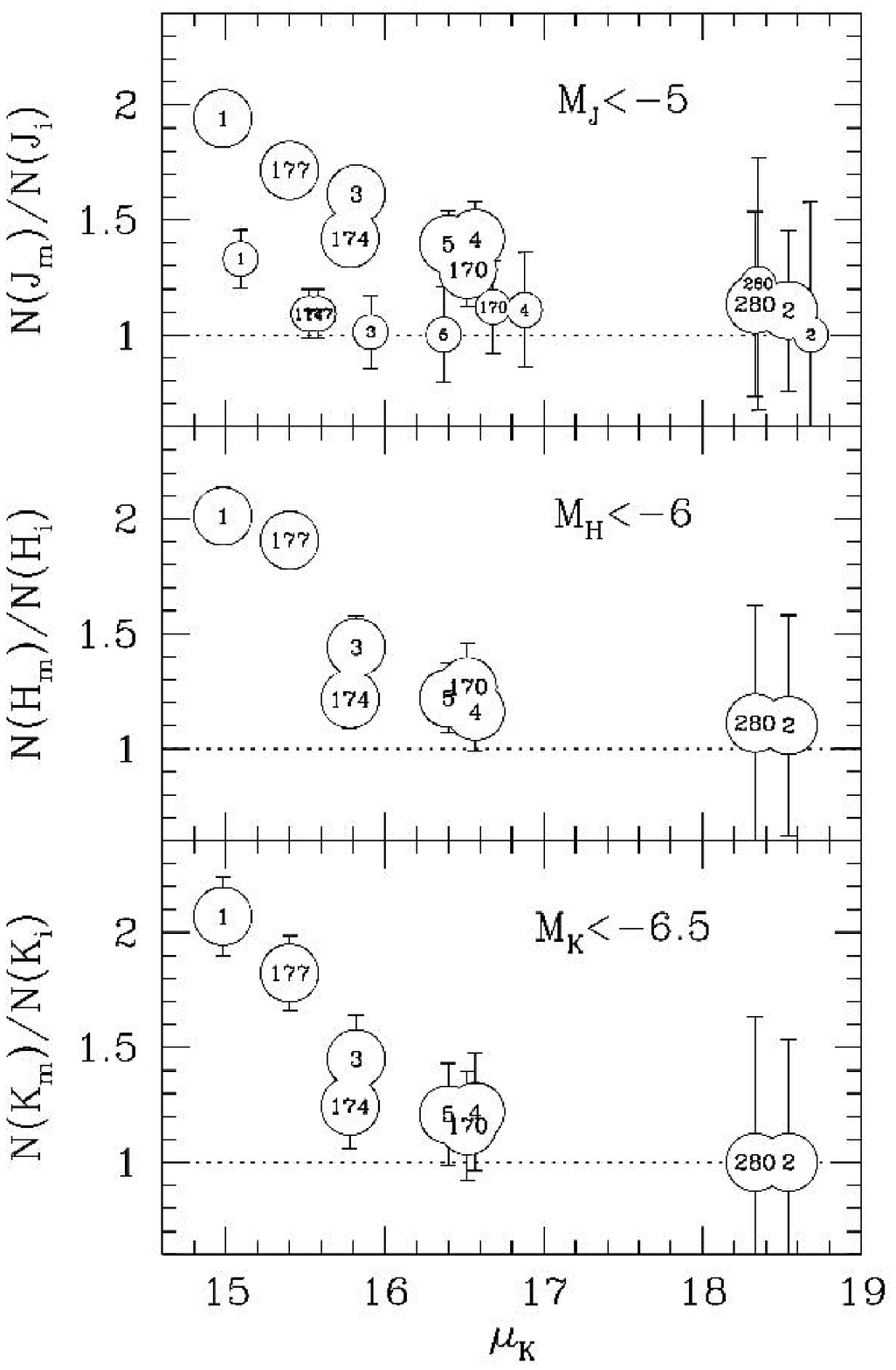} 
\figcaption{
The ratio of the number of bright stars measured in each simulated field
to the number of bright stars input plotted as a function of field
surface brightness (mag/arcsec$^2$).  The top panel shows this ratio for
all the $J$-band frames, where we have chosen $M_J<-5$ as the criterion
to be counted as a bright star.  In this plot the NIC1 fields (small
circles) are less affected by blending than their NIC2 counterparts
(large circles).  The middle panel shows the ratio of measured to input
bright stars in all of our $H$-band frames, where we only count stars
with $M_H<-6$, and the bottom panel gives the ratio for the $K$-band
frames, using $M_K<-6.5$.
\label{fig:ratio}}
\end{figure}

Another interesting, and hopefully more robust, quantity to calculate
for each field is the ratio of the number of ``bright'' stars measured
compared to the number of stars input to the same brightness.  This
quantity is very important, for example when using AGB stars to assess
recent star formation.  We plot this ratio as a function of the field
surface brightness in Figure \ref{fig:ratio}.  We chose $M_J<-5$,
$M_H<-6$, and $M_K<-6.5$ as the criterion for a star to be considered
``bright''.  These limits are fairly arbitrary, however if chosen to be
much brighter, then some of the fields will have no stars input that
bright, and if much fainter, some fields will not be complete to that
level, and we will be measuring completeness instead of blending.

Figure \ref{fig:ratio} shows the ratio of measured to input bright stars
for all of our simulated frames.  The large circles represent NIC2
fields and the small circles NIC1 fields.  The lowest surface brightness
fields (F2 \& F280) have nearly equal numbers of measured and input
stars, i.e. minimal blending.  However, as the surface brightness
increases, the points begin to move up off of the dashed line indicating
a ratio of unity.  This upturn is a function of wavelength, but in
general occurs at $\mu_K \sim 17$ mag/arcsec$^2$ for the NIC2
observations, and $\mu_K \sim 15.5$ mag/arcsec$^2$ for the NIC1
observations.  In the simulations of the brightest fields we measure
about twice as many bright stars in NIC2, and 1.25 times as many in NIC1
compared to the number which were actually input into the simulation.

Of course the ratio of the number of measured to input stars is
dependent on exactly where one draws cutoff magnitude.  At brighter
cutoff magnitudes the ratio goes to infinity when there are no stars
input as bright as we measure.  Choosing a fainter cutoff both dilutes
the number of blends, and causes faint blends to be lost due to
incompleteness.

In summary, both Figures \ref{fig:deltamag} and \ref{fig:ratio} show
similar structure, with a sharp increase in blending between $\mu_K \sim
16 - 17$.  Obviously the amount of blending one can withstand depends on
the scientific goals, however one must be very careful interpreting
results from such data.  We are even skeptical of some of our own
measurements of stars just above the tip of the AGB, which only occur in
the most crowded fields where the surface brightness is greater than
$\mu_K \sim 16$ magnitudes arcsecond$^{-2}$ (however see \S
\ref{sec:brightstars}).

\clearpage

\begin{deluxetable}{cccc}
\tablewidth{0pt}
\tablecaption{Numbers of stars}
\tabletypesize{\footnotesize}
\tablehead{
\colhead{}		&
\colhead{}		&
\multicolumn{2}{c}{\underline{\hspace{0.5cm} Recovered \hspace{0.5cm}}} \\
\colhead{Field}		&
\colhead{Input}		&
\colhead{Real}		&
\colhead{Simulation}	}
\startdata
NIC1	&		&	&	\\
F1	&  900000	& 4037	& 3277	\\
F2	&   25000	&  348	&  521	\\
F3	&  400000	& 3102	& 3088	\\
F4	&  150000	& 2208	& 2411	\\
F5	&  250000	& 2323	& 2405	\\
F170	&  300000	& 2154	& 2055	\\
F174	&  950000	& 4847	& 4970	\\
F177	&  950000	& 4861	& 4970	\\
F280	&   50000	&  448	&  619	\\ \hline
NIC2	&		&	&	\\
F1	& 3000000	& 4037	& 3594	\\
F2	&   80000	&  314	&  597	\\
F3	& 1500000	& 3128	& 3498	\\
F4	&  700000	& 2066	& 2598	\\
F5	&  800000	& 2257	& 2555	\\
F170	&  900000	& 1986	& 2496	\\
F174	& 2000000	& 2923	& 3156	\\
F177	& 4000000	& 3221	& 3010	\\
F280	&  110000	&  256	&  487	\\
\enddata
\label{tab:simulations}
\end{deluxetable}

\section{Comparison with Previous Observations} \label{sec:comparison}

\subsection{\citet{Rich1993}} \label{sec:rmg93}

The groundbased observations of \citet[][hereafter RMG93]{Rich1993} are
the foundation of the current work.  Theirs was the first study in M31
which systematically attempted to measure the stellar properties over a
range of galactocentric distances and bulge-to-disk ratios.  For that
reason our NIC2 fields were chosen to be the same as those in RMG93.
RMG93 took their observations on 1992 Aug 30 -- Sept 1 with the Palomar
IR imager on the Hale 5m telescope.  This instrument was outfitted with
a $58 \times 62$ pixel InSb detector with $0.313''$ pixels.  Each of
their $18'' \times 19''$ fields were observed for 75 seconds through
both $J$ and $K$ filters.  Using offsets equal to half the field of
view, they obtained 25 frames that were later assembled into $72.6''
\times 77.6''$ mosaics.  Thus the central $60''$ of their mosaics have
total integration times of 150s in each filter.  The seeing is $\sim
1''$ on these mosaics.

\begin{figure*}
\epsscale{2}
\plotone{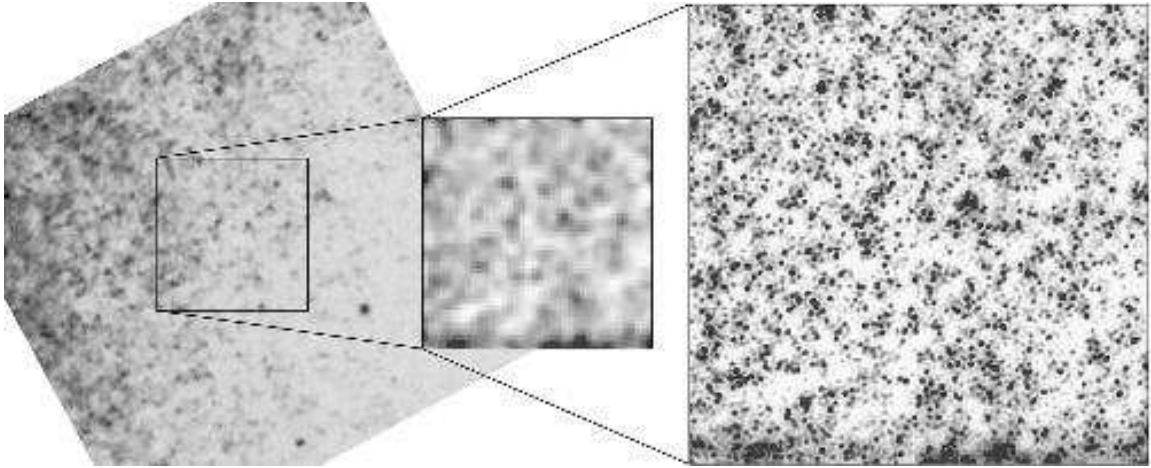} 
\figcaption{
Comparison between groundbased $K$-band image of Field1 (left) and the
NIC2 F222M image (right).  The center image is a rebinned and smoothed
version of the NIC2 image used to help match up the observations.
\label{fig:field1matchup}}
\end{figure*}

The central $\sim 60''$ of Field 1 from RMG93 is shown on the left side
of Figure \ref{fig:field1matchup}.  In order to match up our
corresponding NIC2 image shown in the right side of Figure
\ref{fig:field1matchup}, we first rebinned our image by a factor of 4.1
to go from our $0.0757''$ pixels to the groundbased $0.303''$ pixels,
then smoothed the rebinned image with a Gaussian kernel to match the
$1''$ seeing in the groundbased images.  This rebinned and smoothed
intermediary NICMOS image is shown in the center of Fig
\ref{fig:field1matchup}.

In order to better understand the relationship between the groundbased
photometry and our NICMOS measurements we have performed a star-by-star
comparison of the objects measured by RMG93 in Field 1.  As is obvious
from Figure \ref{fig:field1matchup}, none of the objects seen from the
ground correspond to single stars in the NICMOS image.  However, if we
simply take our brightest measured star nearest the RMG93 object as the
center of the clump which composes their object, we can study the
composition of that clump.

As an example consider RMG93 star 95.  This object is located just above
and right of the center of the groundbased image.  RMG93 measure this
object as $K=15.15$, however the star we match this object with has (the
first entry in Table \ref{tab:star61}) $K=16.63$.  The next star has
$K=16.35$ and lies $0.15''$ away.  If we include all stars within a
radius of $0.5''$ we should get approximately the same amount of flux as
measured by RMG93 viewed through $1''$ seeing.  Table \ref{tab:star61}
lists the NICMOS measured stars and their radius from the assumed center
of the groundbased clump (columns 2 \& 3) and the magnitude of the
running sum of their flux (column 4).  Going down through this table, we
eventually add enough stars to reach the groundbased measurement of
$K=15.15$ at $r \sim 0.52''$.  Of course this radius of equal
measurements varies from star to star, depending on how PSFs were fit to
the groundbased blends.  Using 52 ``stars'' matched with the RMG93
observations, we find this average radius to be $0.47''$ in $K$ and
$0.35''$ in $J$ (see Appendix \ref{app:nic1cal}).  Of the 13 stars with
matches in field 1, the average difference between the RMG93 measurement
and the brightest NICMOS star measured within $0.47''$ is $-1.41$
magnitudes $(\sigma = 0.30)$.

\begin{deluxetable}{cccc}
\tablewidth{0pt}
\tablecaption{RMG93 Star 95 ($K=15.15$)}
\tabletypesize{\footnotesize}
\tablehead{
\colhead{N}		&
\colhead{Distance\tablenotemark{a}}	&
\colhead{$K$}		&
\colhead{Sum}		}
\startdata
  1 &   0.01 &  16.63 &  16.63 \\ 
  2 &   0.15 &  16.35 &  15.73 \\ 
  3 &   0.25 &  17.87 &  15.59 \\ 
  4 &   0.25 &  17.24 &  15.37 \\ 
  5 &   0.44 &  18.68 &  15.32 \\ 
  6 &   0.47 &  18.49 &  15.26 \\ 
  7 &   0.48 &  18.75 &  15.22 \\ 
  8 &   0.51 &  19.47 &  15.20 \\ 
  9 &   0.52 &  18.84 &  15.16 \\ 
 10 &   0.52 &  19.27 &  15.14 \\ 
 11 &   0.57 &  19.47 &  15.12 \\ 
 12 &   0.59 &  17.09 &  14.96 \\ 
\enddata
\tablenotetext{a}{Distance in arcseconds.}
\label{tab:star61}
\end{deluxetable}

\begin{figure}[htb]
\epsscale{1}
\plotone{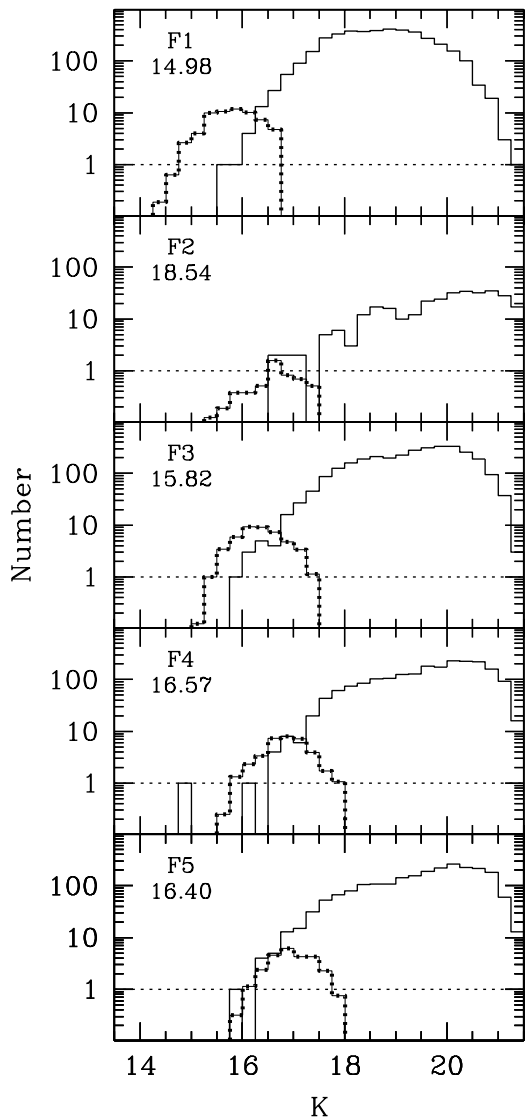} 
\figcaption{
Comparison between our NIC2 measured LFs (solid lines) and the LFs
measured by \citet{Rich1993} (beaded lines) scaled to match the NICMOS
field area.  The $K$-band surface brightness of each field is listed
under the field label in the upper left of each panel.
\label{fig:lfcomprmg93}}
\end{figure}

Figure \ref{fig:lfcomprmg93} shows a comparison between the NICMOS
measured LFs (solid lines) with RMG93's groundbased LFs (beaded lines).
The RMG93 LFs are the measured numbers of stars matched in their $J$ and
$K$-band images placed into 0.25 magnitude bins (column 2 of their Table
8).  We have normalized these LFs by multiplying by 0.063 to compensate
for the larger area of the groundbased images (5634 arcsec$^2$) compared
to the NIC2 area (355 arcsec$^2$).  If the normalized RMG93 LF falls
below a value of one (dashed line) it can be taken as the probability
that NICMOS would find a star that bright if they exist.

Looking at the comparison in Figure \ref{fig:lfcomprmg93} we see that
the amount of disagreement is strongly correlated to the field surface
brightness, which is listed in the upper left of each panel.  Field 1
exhibits the worst case of blending, where RMG93 finds a significant
number of stars a magnitude brighter than we measure.  Field 3 is not as
severe, however the groundbased observations predict that we should find
many more bright stars up to $\sim 0.5$ mag brighter than we see.  The
three lowest surface brightness fields (F2, F4, F5) are roughly
consistent with the NICMOS measurements.  In every field the RMG93 LF
trails off to very bright magnitudes, and even though there is a very
low probability of finding such bright stars in one of our small NIC2
fields, it seems clear that these are most likely just severe cases of
blending.

\subsection{\citet{Rich1991}} \label{sec:rm91}

\citet[][hereafter RM91]{Rich1991} obtained the first infrared color-
magnitude diagram of an M31 bulge field, $3.65'$ from the nucleus and
along the major axis.  They observed $J$ and $K$-band mosaics of nine
non-contiguous $18'' \times 19''$ fields using the Palomar IR imager on
the Hale 5m telescope, yielding a total area of $\sim 3110$ square
arcseconds.

\begin{figure}[htb]
\epsscale{1}
\plotone{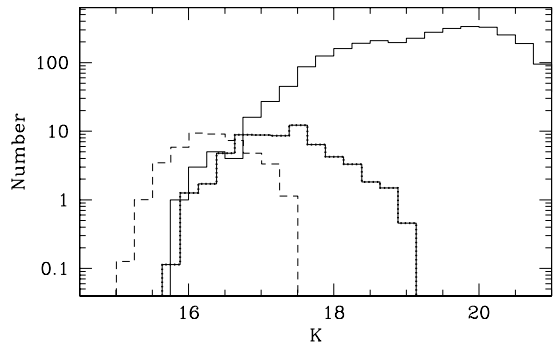} 
\figcaption{
Comparison between our F3 $K$-band luminosity function (solid line) and
the LF measured by \citet{Rich1991} (beaded line).  Also plotted is the
field 3 measurement of \citet{Rich1993}.  We have scaled the groundbased
data to match the NICMOS field area.
\label{fig:rm91}}
\end{figure}

The RM91 field lies close to our F3 field, and we compare the luminosity
functions by taking the RM91 counts from column 4 (CMD) of their Table 3.
We then normalize the RM91 LF by multiplying by the ratio of the NICMOS
field F3 area (355 arcsec$^2$) to the RM91 field area.  The resulting
comparison is shown in Figure \ref{fig:rm91}.

Figure \ref{fig:rm91} shows remarkable agreement between the bright ends
of the NICMOS and RM91 luminosity functions.  Had the M31 bulge
population been normalized to Baade's Window in RM91, the apparent
extended giant branch would have been far less prominent.  However an
accurate normalization is difficult without the faint end completeness
provided by the NICMOS images.

Figure \ref{fig:rm91} also shows a discrepancy between the luminosity
functions of RM91 and RMG93, where the earlier RM91 groundbased data
seems to match the bright end of the NICMOS LF, while the later RMG93
data appears much more affected by blending. 

One possible explanation for this difference seems to lie in the analysis of
the data.  While both datasets were acquired with the same telescope and
instrument, the RM91 data were analyzed on a frame-by-frame basis, while
the RMG93 data were assembled into a large mosaic before analysis.
Therefore the RM91 data retained its original image quality, while the
RMG93 image quality, due to difficulties in perfectly registering the
frames, was reduced to perhaps even worse than that of the worst image.

In light of our simulations, it is still difficult to understand the apparent
agreement between the bright ends of the RM91 and NICMOS luminosity
functions.  Our simulations (for their quoted $0.6''$ seeing) would
still lead to the inescapable conclusion that the measured magnitudes of
the brightest stars in RM91 must still suffer from crowding.  It is also
possible that the images may have been so undersampled ($0.31''$ pixels)
that that the actual seeing was better than quoted, though RM91 state
that 0.6 arcsec seeing was reached in only one of the images.  As the
original frames are not available, we are not able double check the
measurements or to explore the issue in further detail.

\subsection{\citet{Davidge2001}} \label{sec:davidge}

\citet{Davidge2001} has recently obtained $JHK\!s$ images of a bulge field 
$2.6'$ SW of the nucleus of M31 using the 3.6 meter Canada-France-Hawaii
Telescope (CFHT).  With the help of adaptive optics, his images achieve
a FWHM of $0.35''$.  His photometric uncertainties include 0.05
magnitudes in the aperture correction and 0.03 magnitudes in the
zero-point.  The surface brightness of his field (00\h42\m45.1\s,
+41\degr$13' 31.3''$, J2000) is $\mu_K \sim 15.5$ mag/arcsec$^2$ based on
the measurements of \citet{Kent1989} and assuming $(r-K)=2.9$.

\begin{figure}[htb]
\epsscale{1}
\plotone{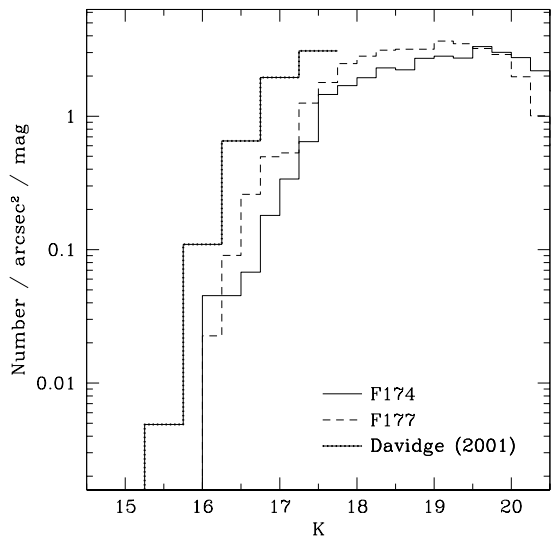} 
\figcaption{
Comparison between the M31 bulge LFs measured by NICMOS: F174 (solid
line) and F177 (dashed line), with that measured by \citet{Davidge2001}
(beaded line).  The surface brightnesses of each field is: 15.8, 15.4, and
15.5 $K$ magnitudes per arcsec$^2$ for F174, F177, and Davidge's field
respectively.
\label{fig:lfcompdavidge}}
\end{figure}

Figure \ref{fig:lfcompdavidge} shows a comparison between the LFs of the
\citet{Davidge2001} field and the two NICMOS fields with bracketing
surface brightnesses: F174 and F177, which have $\mu_K=15.8$ and
$\mu_K=15.4$ mag/arcsec$^2$ respectively.  The Davidge LF should lie
between the two NICMOS LFs, however the Davidge LF is instead shifted
$\sim 0.5$ magnitudes brighter.

Davidge has suggested that the difference between his measurements and
our HST-NICMOS observations \citep{Stephens2001b} is due primarily to
calibration.  In support of this conclusion he cites simulations which
indicate that the effects of blending on his observations are at most
0.1 mag over what would be measured with the NIC2 resolution.  However,
these simulations used simple Gaussian PSFs and included only $\sim
10000$ stars ($<10$ stars/arcsec$^2$).  They show only a few severely
blended stars which he claims are ``easily identifiable''.  In reality
blending is a stochastic phenomenon, involving millions of stars, and
producing a continuum of blending which can be {\it very} difficult to
detect and quantify.

As shown by Figure \ref{fig:nrgbt_reselement}, making observations at
$\mu_K=15.5$ mag/arcsec$^2$ at Davidge's $0.35''$ resolution will give,
on average, $\sim 0.4$ NRGBT stars per resolution element.  Thus in any
given resolution element there is a $(0.4)^2 = 0.16$ probability of
having 2 NRGBT stars, compared to the $(0.07)^2 = 0.049$ probability
with NICMOS.  Thus the most likely cause for the difference between the
\citet{Davidge2001} LF and our NICMOS LFs in Figure
\ref{fig:lfcompdavidge} is indeed due to blending.

As we discussed in Section \ref{sec:simulations}, we are certainly not
claiming the comparison of our observations with \citet{Davidge2001} is
a case of right and wrong; but rather a case of wrong and wrong.  Both
sets of observations are affected by blending, and the stars just above
the AGB tip in our most crowded fields may in fact be artificially
brightened by several tenths of a magnitude.

\section{Bright Stars} \label{sec:brightstars}

We have cautioned that some of the bright stars in our fields may be
blends of fainter stars, however there exists a population of bright
stars which are real.  Inspection of the images shows that these stars
are obvious point sources, and occur over the entire range of M31
stellar densities.  These stars are some of the brightest and bluest
which we have observed, with $16.68 > K > 13.75$ and $0.45 < (J-K) <
0.75$, and are most likely foreground Milky Way stars.  Here we provide
a brief discussion of their properties to ensure that they are not
confused with a population of young M31 bulge stars.

\begin{figure}[htb]
\epsscale{1}
\plotone{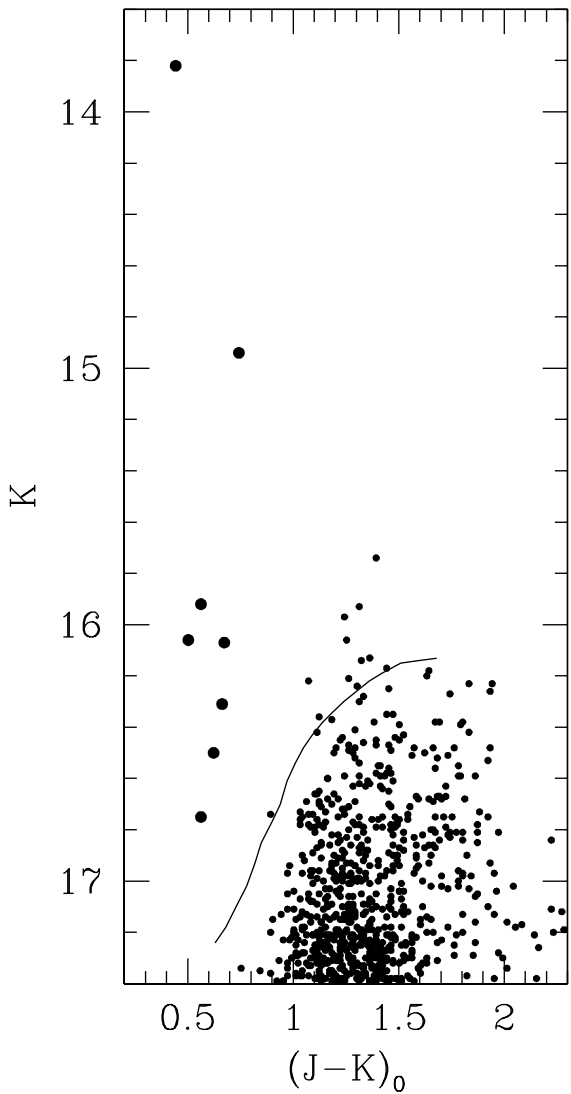} 
\figcaption{
The top end of the combined CMD for all (NIC2) frames.  Note the clear
separation between the red M31 AGB stars and the bluer foreground MW
stars.  The curved line illustrates $M_{bol}=-5$, thus {\it if} these stars
were at the distance of M31, they would all have bolometric magnitudes
over $-5.5$.
\label{fig:brightcmd}}
\end{figure}

The CMD of brightest stars measured in all of the NIC2 frames is shown
in Figure \ref{fig:brightcmd}.  Here the AGB is located between
$1<(J-K)<2$ and extends up to the curved line at $M_{bol}=-5$.  Stars
directly above the AGB are most likely blends of fainter stars.  However
the bright stars bluer than $(J-K)=0.8$ are indeed real stars.  If they
were at the distance of M31 they would all have bolometric magnitudes
brighter than $M_{bol}=-5.5$.

We find a total of 8 foreground stars in the NIC2 fields: 2 in F2, 1 in
F3, 2 in F4, 1 in F5, 1 in F170, and 1 in the G1 field, which has not
been otherwise analyzed in this paper because of the lack of non-cluster
stars.  We find none in the NIC2 F1, F174, F177, or F280 fields.

Searching the NIC1 observations, we use the fact that the faintest
foreground star in the NIC2 group has $J=17.43$.  Thus if we assume
that any star with a NIC1 $J$-band magnitude brighter than $J=17.5$ is
also a MW contaminant, we find 2 in the NIC1 F170 field, and 1 in the
F177 field.  None of the other 8 NIC1 fields (including G1) have any
stars this bright.  These three stars are also clearly separated from
the rest of the NIC1 AGB LF by a $>0.25$ magnitude gap.

\begin{figure}[htb]
\epsscale{1}
\plotone{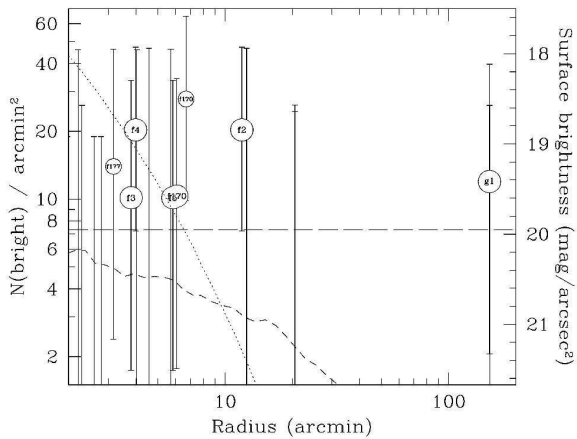} 
\figcaption{
The number of bright foreground stars per square arcminute as a function
of radial distance from the nucleus of M31 (left axis). The long dashed
line at 7.3 shows the mean taken over all fields (11 stars over 1.5
arcmin$^2$).
We have also overplotted the major axis surface brightness profiles of
the bulge (dotted line) and disk (short dashed line) of M31 as measured
by \citet{Kent1989} (right axis).  Here the surface brightness offset is
arbitrary, but the scale is set to match the number counts, so that if
the bright stars were associated with a population in M31, they should
follow the radial surface brightness profile of that population.
\label{fig:nbright}}
\end{figure}

The radial distribution of the bright foreground stars is illustrated in
Figure \ref{fig:nbright}.  Here we plot the surface density measured in
each field as a function of radial distance from the center of M31.  The
upper and lower limits are one-sigma confidence intervals, calculated
using the small-number approximation formulae of \citet{Gehrels1986}.
The distribution shows no trend with radius, and is certainly not
correlated with either of the steeply dropping surface brightness
profiles of M31's bulge or disk, illustrated by the dotted and
short-dashed lines respectively.

In total, we find 8 bright foreground stars in the 10 NIC2 fields.
These fields have a combined area of 0.95 arcmin$^2$, giving a surface
density of 8.4 arcmin$^2$.  Although we don't have color information in
the NIC1 frames, there are three very bright stars with $J<17.5$.  The
total area of the NIC1 fields is 0.56 arcmin$^2$, giving a surface
density of 5.4 stars arcmin$^{-2}$ for these bright stars.  If we assume
that all 11 stars are of the same population, the average surface
density is 7.3 stars arcmin$^{-2}$, which is shown by a long-dashed line
in Figure \ref{fig:nbright}.

%

A comparison with the estimated number of field stars by
\citet{Ratnatunga1985} reveals that our measurement of 7.3 stars
arcmin$^{-2}$ is extremely high.  The measured mean color of $(J-K)=0.6$
corresponds to $(V-K) \simeq 2.2$ for either dwarfs or giants, which
means that these stars most likely have $18.88 > V > 15.95$.  However,
\citet{Ratnatunga1985} predict $\sim 0.8$ field stars / arcmin$^2$
between $V=15$ and $V=19$ toward M31.  Given the combined area of our
NIC1 and NIC2 images (1.5 arcmin$^2$), we should have found
approximately {\em one} field star, not 11.

Unfortunately there have been few surveys of bright field stars toward
M31 to verify the model predictions.  \citet{Ferguson2002} recently
performed a large scale survey of M31 looking for substructure in the
halo and disk of M31.  Integrating over all their magnitudes, $18 <${\it
i}$< 23$, which correspond to roughly $19 < V < 24$, they estimate
13,000 - 20,000 Galactic foreground stars deg$^{-2}$ or between 3.6 -
5.56 stars arcmin$^{-2}$.  This is very close to the prediction of
\citet{Ratnatunga1985} for their magnitude range ($\sim 3.5$
/arcmin$^2$), but the stars we have observed mostly have $V < 19$, and
thus are too bright to be included in their survey.







Thus while the crowded fields have many blends at $M_{bol} \sim -5$, it
is clear that the brightest and bluest stars, with $K < 16.8$ and
$(J-K)<0.8$, are real.  We find 11 of these stars which is over a factor
of 10 greater than what is predicted by Galactic models.  However it
seems very unlikely that these stars are associated with M31 as their
surface density does not scale with the surface brightness of M31, and
we find one as far out as the globular cluster G1, 34 kpc from the
center of M31.

\section{Discussion \& Conclusions} \label{sec:conclusions}

We have analyzed the stellar populations of M31 using 9 sets of adjacent
HST NIC1 and NIC2 fields, with distances ranging from $2'$ to $20'$ from
the nucleus.  These observations are the highest resolution IR
measurements to date, and provide some of the tightest constraints on the
maximum luminosities of stars in the bulge of M31.

Analytic estimates of the effects of blending on our observations, where
we calculate the number of RGB stars within 1 magnitude of the RGB tip
as a function of position in M31, indicate that simulations are required
to accurately interpret our observations.  We thus perform extensive
simulations of each of our NICMOS fields following the procedures of
\citet{Stephens2001a}.  These simulations show that for the most crowded
fields we can expect the brightening due to blending to be as high as
0.75 magnitudes in $J$ and 0.55 magnitudes in $K$.  They also show that
the ratio of measured to input bright stars is a strong function of the
field surface brightness.  In the highest surface brightness field we
measure about 25\% more bright stars than were input with NIC1, and
about twice as many with NIC2.

All of the bulge luminosity functions are consistent with a single,
uniform bulge population.  This is based on the observation that the
small differences we see are correlated with surface brightness, and
that the simulations predict similar differences.  We note however, that
our simulations only verify that a single LF combined with blending can
produce the observed field-to-field differences.  Thus our observations
are {\em consistent} with a single bulge LF, but without higher
resolution data, small field to field differences cannot be ruled out.

The tip of the RGB in M31 is clearly visible at $M_{bol} \sim -3.8$, and
the tip of the bulge AGB extends to $M_K \sim -8$.  This AGB peak
luminosity is significantly fainter than previously claimed.  A
comparison with the measurements of \citet{Rich1993}, which guided our
choice of the 5 pointed observations, indicates that their brightest
stars are most likely severe cases of blending.  In a comparison of our
F174 and F177 LFs with the recently observed bulge field of
\citet{Davidge2001}, we find that the $\sim 0.5$ magnitude difference
between his and our LFs is due entirely to blending in his lower
resolution observations, rather than a calibration error as claimed by
Davidge.

We also find an unusually high number of bright blueish stars in our
fields.  In all 20 (NIC1 + NIC2) fields we find 11 stars which are
uncorrelated with the surface brightness distribution of M31, and appear
to be foreground Milky Way stars.  However, the implied surface density
of 7.3 arcmin$^{-2}$ is over a factor of 10 higher than is predicted by
Galactic models.

\acknowledgements
Support for this work was provided by NASA through grants GO-7826 and
GO-7876 from the Space Telescope Science Institute. 
Thanks to Peter Stetson for supplying, and helping us with his {\sc
allframe} photometry package.  
JAF thanks Dr. Sean Solomon for providing Visiting Investigator
privileges at DTM/CIW.
Also, many thanks to Sergio Ortolani for giving a very helpful referee's
report.

\appendix

\section{NIC1 Transformation} \label{app:nic1cal}

In order to compare observations made with the two different NICMOS
cameras with each other and with groundbased observations we must first
convert all measurements to a common photometric system.  The
transformation of NIC2 to the groundbased CIT/CTIO system has already
been calculated and published \citep{Stephens2000}.  However, NIC1 lacks
a formal transformation to any groundbased system.

As a first attempt to transform NIC1 to a groundbased photometric
system, we applied the NIC2 transformation from \citet{Stephens2000}.
Using their calibration keywords, and assuming $(J-K)=1$ for the color
term, we applied the corresponding offset to all of our NIC1 photometry.
However, this transformation yielded large discrepancies (up to $\sim
0.5$ magnitudes) between the luminosity functions measured in
corresponding NIC1 and NIC2 field pairs, such that the NIC1 fields
appeared too faint.  A comparison between the STScI calibrated NIC1 and
NIC2 F110W luminosity functions shows that the lower surface brightness
fields should show nearly perfect agreement, while the more crowded
fields should show discrepancies of no more than $\sim 0.25$ magnitudes.

As an alternative method to transform NIC1 to a groundbased system, we
considered the observations of \citet{Rich1993}.  Our NIC2 observations
were chosen to be centered on the RMG93 fields, and since the NIC1 and
NIC2 focal planes are so close together ($17.5''$ between field edges)
and NICMOS was rotated $\sim 45$ degrees from North when we took most of
our images, NIC1 falls on the lower left (SE) corner of the RMG93 images
(see Figure \ref{fig:sbmap}).  By rebinning and smoothing our NIC1
images we were able to match up 8 ``stars'' with RMG93.  Of course these
aren't really single stars, but rather clumps of many stars.  However,
by estimating how many stars are in the clumps measured by RMG93, we
were able to use their observations to transform ours to the CIT/CTIO
system.

Going back to our NIC2 observations, whose calibration we trust, we
matched up 52 ``stars'' with RMG93.  Using the $J$-band observations, we
determined that we can make NICMOS agree ($\pm 0.3$ magnitudes) with
RMG93's measurements if we sum up all the NICMOS measured stars within a
$0.35''$ radius around what we estimate to be the center of the RMG93's
``stars''.

To determine the NIC1 transformation, we first calibrate our photometry
using the most recent header keywords listed in the NICMOS Data Handbook
v5.0 \citep{Dickinson2002} namely: {\sc photfnu} = 2.358E-6 Jy sec / DN,
and {\sc fnuvega} = 1773.7 Jy. We then sum the flux of all NICMOS stars
measured within $0.35''$ of the RMG93 centroids.  The resulting
difference between the NIC1 magnitudes and RMG93 is $0.36 \pm 0.17$
magnitudes using all 8 ``stars'', or $0.42 \pm 0.09$ magnitudes using a
sigma-rejected sample of 7.

In summary, we apply a $-0.42$ magnitude offset to our NIC1 F110W
magnitudes to approximately transform them to the $J$-band of the
groundbased CIT/CTIO system.  This is in contrast to the NIC2 photometry
which was transformed using the equations of \citet{Stephens2000}, which
includes a color term in each band.

\end{document}